\documentclass[aps,prx, superscriptaddress,twocolumn,longbibliography]{revtex4-2}

\usepackage{physics}
\usepackage[utf8]{inputenc}
\usepackage{graphicx}
\usepackage[colorlinks=true,linkcolor=black,citecolor=blue,urlcolor=blue]{hyperref}
\usepackage{amsbsy}
\usepackage{xcolor}
\usepackage{amsmath}

\begin{document}

\title{Generation of massively entangled bright states of light during harmonic generation in resonant media} 
\author{Sili Yi}
\affiliation{Max Born Institute, Max-Born Stra{\ss}e 2A, D-12489 Berlin, Germany}
\author{Nikolai D. Klimkin}
\affiliation{Max Born Institute, Max-Born Stra{\ss}e 2A, D-12489 Berlin, Germany}
\author{Graham Gardiner Brown}
\affiliation{Max Born Institute, Max-Born Stra{\ss}e 2A, D-12489 Berlin, Germany}
\author{Olga Smirnova}
\affiliation{Max Born Institute, Max-Born Stra{\ss}e 2A, D-12489 Berlin, Germany}
\affiliation{Department of Physics, Technical University Berlin,  10623 Berlin, Germany}
\affiliation{Technion – Israel Institute of Technology, 3200003 Haifa, Israel}
\author{Serguei Patchkovskii}
\affiliation{Max Born Institute, Max-Born Stra{\ss}e 2A, D-12489 Berlin, Germany}
\author{Ihar Babushkin}
\affiliation{Max Born Institute, Max-Born Stra{\ss}e 2A, D-12489 Berlin, Germany}
\affiliation{Cluster of Excellence PhoenixD (Photonics, Optics, and
  Engineering - Innovation Across Disciplines), Welfengarten 1, 30167
  Hannover, Germany}
\affiliation{Institute for Quantum Optics, Leibniz University of Hannover, Germany}
\author{Misha Ivanov}
\affiliation{Max Born Institute, Max-Born Stra{\ss}e 2A, D-12489 Berlin, Germany}
\affiliation{Technion – Israel Institute of Technology, 3200003 Haifa, Israel}
\affiliation{Department of Physics, Humboldt University, Newtonstra{\ss}e 15, D-12489 Berlin, Germany}
\affiliation{Blackett Laboratory, Imperial College London, South Kensington Campus, SW7 2AZ London, United Kingdom}

\begin{abstract}
    At the fundamental level, full description of light-matter interaction requires quantum treatment of both matter and light. However, for standard light sources generating intense laser pulses carrying quadrillions of photons in a coherent state, the classical description of light during intense laser-matter interaction has been expected to be adequate.  
    Here we show how nonlinear optical response of matter can be controlled to generate dramatic deviations from this standard picture, including generation of several squeezed and entangled harmonics of the incident laser light. 
    In particular, such
    non-trivial quantum states of harmonics are generated as soon as one of the harmonics induces a transition between  
    different laser-dressed states of the material system. Such transitions generate an entangled light-matter wavefunction, which can generate quantum states of harmonics even in the absence of a quantum driving field or material correlations. In turn, entanglement of the material system with a single harmonic generates and controls entanglement between different harmonics. Hence, nonlinear media that are near-resonant with at least one of the harmonics appear to be 
    quite attractive for controlled generation of massively entangled quantum states of light. Our analysis opens remarkable opportunities at the interface of attosecond physics and quantum optics, with implications for quantum information science.   
    
\end{abstract}

\maketitle

\section{Introduction}

Quantum nature of light is central to our understanding of the
micro-world, with the quantum description of the photo-electric effect
ushering the new era in physics at the turn of the XX-th century.
However, quantum description of light is typically used in the regime
where just a few, or a few tens, of photons interact with quantum
matter. In contrast, when intense incident light carries quadrillions
of photons, counting each one of them individually is hardly expected
to matter. This is why, until recently, high harmonic generation in strong laser fields, both in gases and in solids (see e.g. \cite{ferray1988multiple,drescher01,paul01,
sukiasyan2010exchange,Frausz2009,ghimire11,krausz14,vampa14,liu17,ghimire19,goulielmakis22}) has been
considered in the context of classical optics.

Recent observation \cite{lewenstein2021generation} of the Schrödinger cat-type (so-called kitten) states and related statistics \cite{tsatrafyllis17} of just such an intense light upon nonlinear-optical interaction with an atomic gas has upended this conventional wisdom. Granted, the observed Schrödinger kitten states have only emerged upon conditioning the observation of the transmitted driving field on detecting the highly nonlinear optical response to it. Nevertheless,  Ref.\cite{tsatrafyllis17,gorlach20,lewenstein2021generation}, followed by \cite{rivera-dean22,rivera-dean22a,maxwell22,stammer22,stammer22a,Lewenstein2023,bhattacharya23,lemieux2024photon,theidel2024evidence,rasputnyi2024high}, 
have opened an exciting possibility of using high harmonic emission to generate bright quantum states of light by measuring the transmitted laser field together with generated harmonics. 

Hindsight is always 20/20. Looking back, we can now identify several important first steps along this exciting new road. Early theoretical approaches \cite{sundaram90,gauthey95}, which started the formal mathematical treatment of high harmonic generation using second quantization formalism, showed how one can bring the fully quantum picture of the process towards the semi-classical limit, and what assumptions should be made to get there. The semi-classical prejudice started to change with Refs.\cite{gonoskov2016quantum, tsatrafyllis17}. These works posed the key new question: does the quantum nature of light manifest in the properties of the incident laser field after the interaction, especially when correlated to the generated harmonics? The positive answer to this question has steered the quest in the right direction.

Ref.\cite{gorlach20} suggested that nontrivial (albeit extremely weak in Ref.\cite{gorlach20})  quantum features might appear in the properties of the generated light due to interference of multiple quantum pathways generating nonlinear optical response. Quantum properties of the generated light can be further amplified if the incident light interacts with a correlated quantum state of a material system, mapping material correlations onto the generated light \cite{gorlach20,pizzi23,tzallas23,lange2023electron}.

This theoretical optimism is now supported by experimental advances: the promise of the early steps \cite{tsatrafyllis17,lewenstein2021generation} has now been strengthened by Refs.\cite{lemieux2024photon,theidel2024evidence,rasputnyi2024high,lamprou2023nonlinear}. In particular, while the initial experimental route \cite{tsatrafyllis17,lewenstein2021generation} took advantage of a post-selection procedure, the experiment \cite{theidel2024evidence} demonstrated distinctly quantum features of the generated harmonics without relying on it. 

There is, of course, a different route to generating bright quantum states of light via nonlinear optical response: one can use intense driving field already in a quantum state \cite{Spasibko2017,even2023photon, Kaminer2023,rasputnyi2024high}, or combine classical and quantum drivers \cite{lemieux2024photon}. As the quantum properties of the incident light should affect quantum properties of the generated harmonics, the latter should emerge in nontrivial quantum states. The technological breakthrough in generating the so-called bright squeezed vacuum states of light \cite{iskhakov2009generation,iskhakov2012superbunched} with a sufficient number of photons has enabled generation of several harmonics of this light \cite{Spasibko2017,rasputnyi2024high}. 
In this context, we also note the 
recent proposal \cite{sloan2023entangling} to extend the parametric down conversion sources into the UV and XUV range 
by replacing the traditional single high-energy 
pump photon with many low-energy ones.

Here we introduce a completely different avenue. We start with a standard laser pulse in a coherent state, and a quantum material system in a simple, uncorrelated ground state. To generate harmonics in a non-trivial quantum state, we rely only on the excitation of the material system by the generated harmonic light, a process often ignored in the description of high harmonic generation. 
Unsurprisingly, once such excitations are included, the light-matter state becomes entangled. This entanglement is central to what follows. 

First, perhaps unexpectedly, all harmonics correlated to the new excited state of the material system become entangled with each other. Second, and most importantly, control over the excitations of the material system gives access to flexible control over the generated entangled states of light across multiple octaves. Third, such control over excitations in the material system needs to be neither complicated nor sophisticated: it can make use of the ubiquitous Stark shifts induced by the classical laser field, and be as straightforward as modulating the intensity envelope of the driving classical field. 

Our analysis shows that generation of non-trivial quantum states of harmonics does not have to be an exception: it can occur whenever the material system undergoes non-adiabatic transitions between its laser-dressed states. Thus, one does not need to rely on pre-exciting strongly correlated material states \cite{pizzi23} or on
using non-classical incident light \cite{Spasibko2017,even2023photon,Kaminer2023,rasputnyi2024high}. In fact, our results offer one possible interpretation of the 
extraordinary experimental findings of \cite{theidel2024evidence}.

Moreover, our analysis also suggests that post-selection of the generated light  \cite{tsatrafyllis17,lewenstein2021generation,tzallas23}, can be significantly helped by shaping the transverse modes of the generated harmonics, as discussed below. 

The focus on a strong feedback of the generated light onto the material system, and on the quantum features of light induced by such a feedback, is the key difference of our approach from others. In this context, the weakness of quantum features found in \cite{gorlach20} is associated with the lack of such a feedback.

We begin our analysis by first identifying the conditions that preclude generation of nontrivial quantum states of multiple harmonics. This helps us identify the key conditions for generating highly non-classical states of harmonics, starting with a purely classical incident light and a simple material system residing in an uncorrelated ground state. We then discuss one route to implementing these conditions, which to a theorist's eye appears relatively straightforward.

\section{Prerequisites for generating quantum states of  harmonic light}

The equation describing the interaction of light with a quantum system is (atomic units are used throughout): 
\begin{eqnarray}
i\frac{\partial}{\partial t}\ket{\Psi}=\hat{H}\Psi=
\left[
\hat{H}_{_A}+\hat{\textbf{x}}\hat{\textbf{F}}+\hat{H}_{_F}
\right] \ket{\Psi},
\label{eq:FullTDSE_1}	
\end{eqnarray}
where $\hat{H}_{_A}$ is the Hamiltonian of the material system, $\hat{H}_F$ is the Hamiltonian of the quantum field, and $\hat{\textbf{x}}\hat{\textbf{F}}$ describes their coupling in the dipole approximation, with $\hat{\textbf{x}}$ representing the negative of the dipole operator; $\hat{\textbf{F}}$ is the electric field operator acting on all light modes. 
Here, for clarity,
we shall restrict the mode frequencies to $\omega_n=n\omega$, with
$\omega$ the incident frequency. Ignoring harmonic line-widths implies that 
sufficiently long driving pulses are assumed and that 
the bandwidth of the harmonic lines is much more 
narrow than their central wavelengths.

The wavefunction starts in the product state
$\ket{\Psi_0}=\ket{g}\ket{\alpha_{\omega}}$, with $\ket{g}$ the field-free ground state of the material system and $\ket{\alpha_{\omega}}$ the initial coherent state at the fundamental frequency $\omega$.

What happens if the full wavefunction $\ket{\Psi(t)}$ remains, throughout the whole evolution, in the product state of the material system $\ket{\psi_g(t)}$ and of the state of light $\ket{\chi_{_F}(t)}$, 
$\ket{\Psi(t)}=\ket{\psi_g(t)} \ket{\chi_{_F} (t)}$? Here the dressed quantum state of the material system $\ket{\psi_g(t)}$ has evolved from the field-free ground state and can include all states: ground, bound, and continuum; its final overlap with the ground state might even be negligible.  

Substituting the ansatz $\ket{\Psi(t)}=\ket{\psi_g(t)} \ket{\chi_{_F}(t)}$ into the Eq.(\ref{eq:FullTDSE_1}) yields the following equation for the quantum light field,
\begin{eqnarray}
    i\ket{\dot{\chi}_{_F} (t)}=&& \left[ 
    \hat{H}_{_F}+\textbf{x}_{gg}(t)\hat{\textbf{F}}+\Delta E_g(t)
    \right] \ket{\chi_{_F} (t)},
\label{eq:TDSE_Chi_F}
\end{eqnarray}
where $\textbf{x}_{gg}(t)=\bra{\psi_g (t)} \hat{\textbf{x}}\ket{\psi_g(t)}$
is the dipole induced in the material system; the time-dependent energy shift in the material system
$\Delta E_g(t)=\bra{\psi_g(t)}\hat H_A-i\frac{\partial}{\partial t}\ket{\psi_g(t)}$ has no influence on the quantum state of light $|\chi_F\rangle$, which starts in the product of vacuum states for each harmonic. 
A similar equation is obtained for the material system.

For each frequency $n\omega$, Eq.(\ref{eq:TDSE_Chi_F}) describes a harmonic oscillator 
$\hat{H}_{_F,_n}$  starting in its ground state and driven by the time-dependent force $\textbf{x}_{gg}(t)$. No matter how complex $\textbf{x}_{gg}(t)$ is, the linear coupling $\textbf{x}_{gg}(t)\hat{\textbf{F}}$ can only shift the initial vacuum state $\ket{\alpha_n=0}$ to a coherent state $\ket{\alpha_n(t)}$ with a higher average photon number. The factorized nature of the initial state $\ket{\chi_{_F}}$
of the harmonics, which starts as a product state of
individual harmonics in their vacuum states, is preserved, as well as the maximally classical nature of the generated harmonics. What are the possible escape routes from this "classical corner"?

Mathematically, one route is to turn the 
linear Eq.(\ref{eq:TDSE_Chi_F}) for the quantum state of 
a generated harmonic $n$ into nonlinear. That is,   
the material response $\textbf{x}_{gg}(t)$ entering 
Eq.(\ref{eq:TDSE_Chi_F}) for some harmonic $n$ should 
be sensitive to the state $\ket{\chi_{_F,_n}}$ of this harmonic. This removes the curse of limiting the evolution of $\ket{\chi_{_F,_n}}$ 
to a single coherent state.

Physically, the dependence of the material response on the generated 
harmonics implies that they can induce transitions in the material system and change its nonlinear response. Thus,  the quantum states of matter and the state of light become correlated.
The way to incorporate such correlation is to eschew the factorized ansatz $\ket{\Psi(t)}=\ket{\psi_g(t)} \ket{\chi_{_F}(t)}$, which brought us into the unwanted "classical corner" in the first place, explicitly providing for the possibility of light-matter entanglement. Quantum correlations and feedback between the generated nonlinear optical response and the state of the material system are the basis for generating nontrivial quantum states of all harmonics. 

To emphasize the lack of need for quantum light at the input, we shall replace the field operator at the fundamental frequency $\omega$ with the classical field
$\textbf{F}_{\rm cl}(t)=\bra{\alpha_1(t)}\hat{\textbf{F}}_{\rm 1}\ket{\alpha_1(t)}$,
approximating the full Hamiltonian as 
\begin{eqnarray}
&&\hat{\textbf{H}}\simeq \hat{H}_{_A}+\hat{\textbf{x}}\textbf{F}_{\rm cl}(t)+\hat{\textbf{x}}\sum_{n\neq 1}
    \hat{\textbf{F}}_n+
    \sum_{n\neq 1}\hat{H}_{_F,_n},    
    \nonumber \\
    &&\textbf{F}_{\rm cl}(t)=\bra{\alpha_1(t)}\hat{\textbf{F}}_{\rm 1}\ket{\alpha_1(t)}.
    \label{eq:Hamiltonian1}
\end{eqnarray}
Such approximation is by no means necessary, as 
the general analysis is nearly as straightforward as that developed 
below. 
However, this approximation 
allows us to bring to the fore the key role of 
harmonic light-matter entanglement, which
leads to the generation of quantum light at the output without any quantum light at the input.

\begin{figure*}[htp]
\centering
\includegraphics[width = 0.7\linewidth]{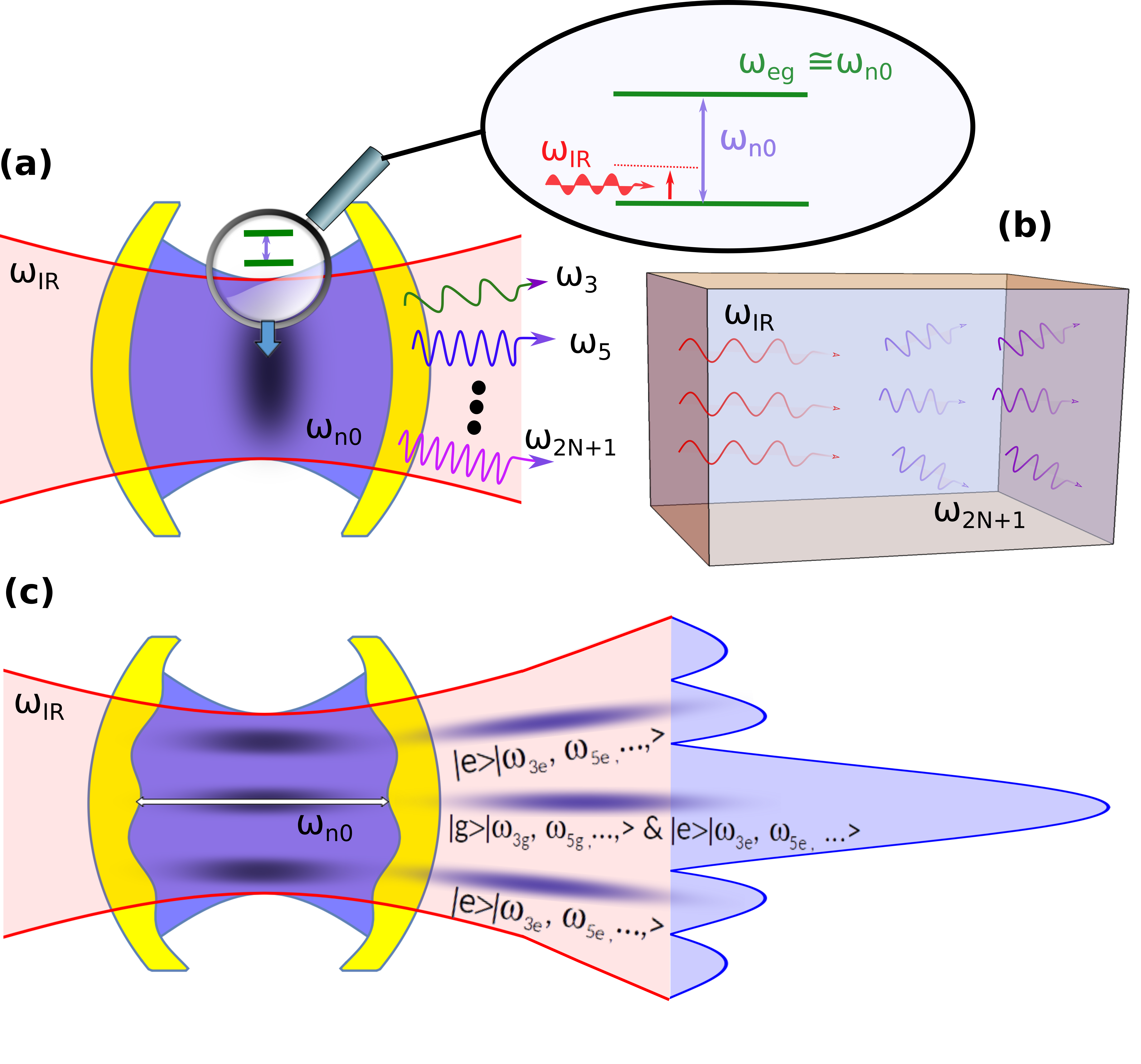}
\caption{
Some schemes for generating harmonics in quantum states. (a) Use of a cavity. Resonance between two laser-dressed states of a quantum system, driven by a low-frequency laser field, and a generated harmonic entangles the material system with the generated light. This also leads to massive entanglement between all harmonics generated upon the excitation. The resonance is enhanced by tuning the resonant cavity mode to the resonant harmonic. (b) Instead of the cavity, one can rely on the absorption of some of the generated harmonics in a sufficiently dense medium, whether solid or sufficiently gas. (c) Transverse shaping of the resonant cavity mode separates in the far field the highly non-classical components of the generated harmonics, correlated to the excited state, 
from the (less interesting) harmonics correlated to the ground state. 
} 
\label{Figure1}
\end{figure*}

How can we enhance the 
feedback of the generated harmonic light on the material system that generates it? Given the relative weakness of the generated harmonics, one possible route is to ensure a one-photon resonance between some  harmonic $n_0\omega$ and the transition between the ground
$\ket{g(t)}$ and some excited $\ket{e(t)}$ dressed state of the material system. 
Importantly, this one-photon resonance does not have to be present at the zero fundamental field. On the contrary, we shall see below that it is beneficial to have such a resonance induced by the Stark shifts of the ground and excited states. 

To further increase the light-matter coupling, one can add a cavity resonant with $n_0\omega$, as shown in Fig.1(a). Fig. 1(b) shows a complementary option of an optically thick medium for some harmonics, which ensures that these harmonics are invested in building the correlation between the state of the material system and the state of the generated light.
Fig.1(c) shows that one can additionally shape the transverse mode of the generated resonant harmonic, separating in the far field all harmonics correlated to the excited state of the driven material system from those correlated to its ground state. 

We shall see that resonant absorption of a \textit{single} harmonic affects the quantum state of all other harmonics: a single resonant transition is sufficient to dramatically affect the quantum properties of \textit{all} generated light fields.

\section{General formalism}

One-photon resonance with a particular harmonic means that there is also an 
$n_0$-photon resonance with the fundamental field. Such multiphoton resonances can already be included in the definition of the laser-driven ground state
$\ket{\psi_g(t)}$: we define $\ket{\psi_g(t)}$ as the solution of the time-dependent Schrödinger equation (TDSE) in the classical incident field $\textbf{F}_{\rm cl}(t)$, for the system starting in the ground state, 
\begin{eqnarray}
&&    i\frac{\partial}{\partial t}\ket{\psi_g(t)}=
    \left[
    \hat{H}_{_A}+\hat{\textbf{x}}\textbf{F}_{\rm cl}(t)
    \right]\ket{\psi_g(t)},
    \nonumber
    \\
&&     \ket{\psi_g(t_i)}=\ket{g}.
\label{eq:ClassicalTDSE_g}
\end{eqnarray}
We will refer to this state as the "classically dressed" ground state.
One-photon absorption of the harmonic $n_0\omega$
occurs between this state and the "classically dressed" excited state
$\ket{\psi_{e}(t)}$ defined as the solution of the standard semi-classical TDSE 
\begin{eqnarray}
&&  i\frac{\partial}{\partial t}\ket{\psi_e(t)}=
    \left[
    \hat{H}_{_A}+\hat{\textbf{x}}\textbf{F}_{\rm cl}(t)
    \right]
    \ket{\psi_e(t)},
    \nonumber
    \\
&&     \ket{\psi_e(t_i)}=\ket{e}.
\label{eq:ClassicalTDSE_e}
\end{eqnarray}
with the initial condition set to the 
excited field-free state $\ket{e}$. All
other excited dressed states of the system are defined
in the same manner.
Note that these dressed states fully incorporate the incident field, remain orthogonal to each other, and offer a perfect basis to incorporate effects of the harmonics on the material system. 

We can now write the full light-matter wavefunction as:
\begin{eqnarray}
\ket{\Psi(t)}=\ket{\psi_g(t)} \ket{\chi_g(t)}+ \sum_e
\ket{\psi_e(t)} \ket{\chi_e(t)},
\label{eq:FullWF_1}	
\end{eqnarray}
where $\ket{\chi_g(t)}, \ket{\chi_e(t)}$ are the states of light correlated to the classically dressed states of the material system. As such, Eq.(\ref{eq:FullWF_1}) is completely general, stating that each dressed (Floquet) state $\ket{\psi_e(t)}$ of the material system is correlated to some yet unknown state of light $\ket{\chi_e(t)}$.  In the semi-classical description of light 
$\ket{\chi_e(t)} = C_e(t) \ket{\chi_g(t)}$ and $C_e(t)$ are 
the excitation amplitudes of the dressed states $\ket{e(t)}$. 
The norms of $\ket{\chi_g(t)}$, $ \ket{\chi_e(t)}$ reflect 
the probability of the material system to populate
the corresponding dressed states $\ket{g(t), e(t)}$, just like
$|C_e(t)|^2, |C_g(t)|^2$ would in the semi-classical case. 
In the semiclassical regime, where $\ket{\chi_e(t)} = C_e(t) \ket{\chi_g(t)}$, no entanglement between light and matter is present. Our aim is to identify conditions where one could expect strong deviations from this regime.

Substituting Eq.(\ref{eq:FullWF_1}) into the full Schrödinger equation yields coupled equations for the light fields
correlated to the strongly driven states of the material system,
\begin{eqnarray}
    i\ket{\dot{\chi}_{g}}= \hat{H}_{_F}\ket{\chi_g}+\textbf{x}_{gg}(t)\hat{\textbf{F}}\ket{\chi_g}+
    \sum_e\textbf{x}_{ge}(t)\hat{\textbf{F}}\ket{\chi_e},
\label{eq:TDSE_Full_Chi_g}
\end{eqnarray}
\begin{eqnarray}
    i\ket{\dot{\chi}_{e}}= \hat{H}_{_F}\ket{\chi_e}+\sum_{e'}\textbf{x}_{ee'}(t)\hat{\textbf{F}}\ket{\chi_{e'}} +
    \textbf{x}_{eg}(t)\hat{\textbf{F}}\ket{\chi_g},
\label{eq:TDSE_Full_Chi_e}
\end{eqnarray}
where $\hat{\textbf{F}}=\sum_{n\neq 1}\hat{\textbf{F}}_n$ is the field operator acting on all frequencies except the 
fundamental $n=1$,
\begin{eqnarray}
    \textbf{x}_{gg}(t)=\bra{\psi_g(t)}\hat{\textbf{x}}\ket{\psi_g(t)},
 \nonumber \\
    \textbf{x}_{ee}(t)=\bra{\psi_e(t)}\hat{\textbf{x}}\ket{\psi_e(t)},
\label{eq:x_gg(t)}
\end{eqnarray}
are the laser-induced dipoles in the dressed ground and excited states, 
and 
\begin{eqnarray}
\textbf{x}_{ge}(t)=\bra{\psi_g(t)}\hat{\textbf{x}}\ket{\psi_e(t)},
\nonumber \\
\textbf{x}_{eg}(t)=\bra{\psi_e(t)}\hat{\textbf{x}}\ket{\psi_g(t)},
\nonumber \\
\textbf{x}_{ee'}(t)=\bra{\psi_e(t)}\hat{\textbf{x}}\ket{\psi_{e'}(t)},
\label{eq:x_eg(t)}
\end{eqnarray}
describe the transition dipole matrix elements between the states
dressed by the classical field.

Several comments are now in order. 
First, as we have already mentioned, the dressed atomic states $\ket{\psi_e(t)}$, $\ket{\psi_g(t)}$ incorporate many field-free states, in principle including continuum states. 
Second, our description does not imply any specific harmonic generation
mechanism. All one needs are the time-dependent dipoles obtained from the solution of the TDSE for the classical incident driving field
and the material system starting in all relevant field-free states.
Therein lies the challenge: quantum description of light requires 
one to incorporate all 
essential dressed states, solving as many classical
TDSEs as needed to describe the physics underlying the
generation process.
To make things tractable, 
below we shall limit quantum description of the generated 
light to the harmonics of the driver, tacitly ignoring the 
width of each harmonic. This approximation assumes sufficiently long
driving pulses.

For moderate laser intensities and 
a single pair of resonantly coupled dressed states, including 
only two dressed states in the analysis is another reasonable approximation.
We shall refer to this as  \textit{two-channel approximation}, to  
distinguish it from the much simpler case of a 
pure two-level system. The difference 
is clear: in the two-channel case, each dressed state may involve
infinitely many field-free states of the material system.

Below we shall focus on moderate driving laser intensities and assume that
the generated harmonics are relatively weak. In this case
one can use first-order perturbation theory with respect to
the effect of the generated harmonics on the material system.
In this limit, the terms coupling the states $\ket{e(t)}$ and
$\ket{e'(t)}$ in Eq.(\ref{eq:TDSE_Full_Chi_e}) are omitted. Moreover, we shall also use the two-channel approximation and only keep the ground dressed state and the excited dressed state resonant with it. This does not, however, imply the two-level approximation.
 
In the limit of sufficiently weak  driving 
\begin{eqnarray}
\textbf{x}_{eg}(t)=\bra{\psi_e(t)}\hat{\textbf{x}}\ket{\psi_g(t)}\rightarrow
\textbf{x}_{eg}e^{i\omega_{eg}(t-t_i)},
\label{eq:x_eg(t)_2}
\end{eqnarray}
where $\omega_{eg}=E_e-E_g$ is the transition frequency and 
$t_i$ is some initial moment of time. In stronger fields, these transition dipoles can acquire non-trivial time-dependence, e.g. through the time-dependence of the 
transition frequency $\omega_{eg}(t)=E_e(t)-E_g(t)$ which
incorporates the Stark shifts of both states. 
We shall take advantage of this ubiquitous effect below.

We now take advantage of the interaction picture, defining $|\chi_{g,e}(t)\rangle=e^{-i\hat{H}_{_F} t}|\phi_{g,e}(t)\rangle$.
With this substitution, we have 
\begin{eqnarray}
    i\ket{\dot{\phi}_g (t)}=&\textbf{x}_{gg}(t)\hat{\textbf{F}}_{I}(t)\ket{\phi_g}+
    \textbf{x}_{ge}(t)\hat{\textbf{F}}_{I}(t)\ket{\phi_e},
    \nonumber
    \\
    i\ket{\dot{\phi}_e (t)}=&\textbf{x}_{ee}(t)\hat{\textbf{F}}_{I}(t)\ket{\phi_e}+
    \textbf{x}_{eg}(t)\hat{\textbf{F}}_{I}(t)\ket{\phi_g},
\label{eq:TDSE_Full_InteractionPicture1}
\end{eqnarray}
For each frequency $n\omega$, the time-dependent operator $\hat{\textbf{F}}_{_I}(t)$ is given by the corresponding creation 
and annihilation operators $\hat{a}_{_I,_n}(t)=\hat{a}_ne^{-in\omega t}$.
Note that the definition of $\hat{\textbf{F}}$ in terms of 
$\hat{a}$ and $\hat{a}^{\dagger}$ admits
certain freedom for the relative phase between $\hat{a}$ and $\hat{a}^{\dagger}$ \cite{grynberg2010introduction}.
Indeed, if the fundamental classical field $\textbf{F}_{\rm cl,1}(t)=\langle\alpha_1(t)|\hat{\textbf{F}_1}|\alpha_1(t)\rangle$ 
is set as $\mathbf{ F}_{\rm cl,1}(t)\propto \cos\omega t\propto e^{-i\omega t}+e^{+i\omega t}$, then
in the interaction picture 
\begin{eqnarray}
    \hat{\textbf{F}}_{I,1}(t)=
    \boldsymbol{\mathcal{F}}_1f_1(t)
    \left[
    \hat{a}^{\dagger}_1e^{+i\omega t}+\hat{a}_1e^{-i\omega t}
    \right],
\label{eq:FieldOperatorSP_1}
\end{eqnarray}
where 
$\boldsymbol{\mathcal{F}}_1$ is the vacuum field amplitude for the fundamental. 
Here we used that, for a coherent state $\ket{\alpha(t)}$, 
$\bra{\alpha(t)}\hat{a}\ket{\alpha(t)}=\bra{\alpha}\hat{a}_{_I}(t)\ket{\alpha}=\alpha(t).$

The very slow envelope $f_1(t)$ was added 
to account for the finite pulse length, assuming sufficiently long driving pulse with a
very large number of cycles. We shall use the same convention for the field operators at all other relevant frequencies, with 
$\boldsymbol{\mathcal{F}}_n$ the vacuum field amplitude for the harmonic $n$. 
The relative phase between the fundamental and
the harmonics manifests in the phases of the quantum states for the harmonic modes.
The discussion of the relevant quantization volume required to quantify 
$\boldsymbol{\mathcal{F}}_n$ is presented in Section 4A.

The simplest route to move forward with the coupled equations
Eq.(\ref{eq:TDSE_Full_InteractionPicture1}) is to
use the time-dependent 
perturbation theory. Its zero-order equation describes the light field correlated to the dressed ground state, 
\begin{eqnarray}
    i\ket{\dot{\phi}_{g}^{(0)}(t)}=\textbf{x}_{gg}(t)\hat{\textbf{F}}_{I}(t)\ket{\phi_{g}^{(0)}(t)}
\label{eq:TDSE_Interaction_ZeroOrder}
\end{eqnarray}
with the initial condition being the vacuum state for all harmonics. Its formal solution is
$\ket{\phi_{g}^{(0)}(t)}=\hat{U}^{(0)}_g(t,t_i)\ket{\phi_{g}^{(0)}(t_i)}$
where $\hat{U}^{(0)}_g(t,t_i)$ is the propagator for the homogeneous equation
Eq.(\ref{eq:TDSE_Interaction_ZeroOrder}) with the driving function 
$\textbf{x}_{gg}(t)$. It is a product of propagators for each harmonic,
\begin{eqnarray}\hat{U}^{(0)}_g(t,t_i)=\prod_n\hat{U}^{(0)}_{g,n}(t,t_i).
\label{eq:HomogeneousPropagatorG_n}
\end{eqnarray}
At any time $t$, the quantum state of generated harmonics remains a product state. 
Moreover, for each harmonic $n$ the propagator $\hat{U}^{(0)}_{g,n}(t,t_i)$ simply shifts the initial vacuum state $|\alpha_n(t_i)=0\rangle$
to the coherent state 
$|\alpha_n(t)\rangle$,  $|\alpha_n(t)\rangle=\hat{U}^{(0)}_{g,n}(t,t_i)|\alpha_n(t_i)\rangle$,
where 
\begin{eqnarray}
    \alpha_n(t)=-i 
    \boldsymbol{\mathcal{F}}_n\int_{t_i}^{t} d\tau \textbf{x}_{gg,n}(\tau).
\label{eq:Alpha(t)}
\end{eqnarray}
Here $\textbf{x}_{gg,n}$ are the complex amplitudes of the Fourier components 
of the full dipole response $\textbf{x}_{gg}(t)$ at the frequency $n\omega$.
Note that
both $\textbf{x}_{gg,n}$ and $\alpha_n$ are complex-valued.
The phase of $\alpha_n$
accounts, in particular, for the phase of the harmonic relative to the 
driving field. For low-frequency drivers and at 
modest intensities, before the onset of ionization, this phase is negligible, as harmonics are in phase with the driving field. This contrasts with 
recollision-based harmonics, where the harmonic phase is
significant and order-dependent.
Note also the slow time dependence included in the Fourier components, 
$\mathbf{x}_{gg,n}(\tau)$, which accounts for their
envelopes. Formally, such envelope  is introduced via the gated Fourier transform 
$\mathbf{x}_{gg,n}(\tau)=\int G(t-\tau)\mathbf{x}_{gg}(t)e^{in\omega t}dt$ with a suitable gate $G(t)$, which separates fast (optical) and slow (envelope) time scales, valid for 
sufficiently long pulses.

The zero-order solution is then substituted into the equation for the field correlated to the dressed excited state, yielding 
\begin{eqnarray}
\label{eq:SolutionFirstOrderE}
    &&|\phi_{e}^{(1)}(t)\rangle =
    \\
    &&=-i\int^T_{t_i} dt'  \hat{U}^{(0)}_e(T,t')
    \textbf{x}_{eg}(t')\hat{\textbf{F}}_{I}(t')\hat{U}^{(0)}_g(t',t_i)|\phi_{g}^{(0)}(t_i)\rangle,
    \nonumber 
\end{eqnarray}
where the propagator 
$\hat{U}^{(0)}_e(t,\tau)$ solves the homogeneous equation similar to 
Eq.(\ref{eq:TDSE_Interaction_ZeroOrder}) but 
for the laser-induced dipole generated by  the material system in the dressed excited state,  
$\textbf{x}_{ee}(t)$.

\begin{figure*}[htp]
\centering
\includegraphics[width = 1\linewidth]{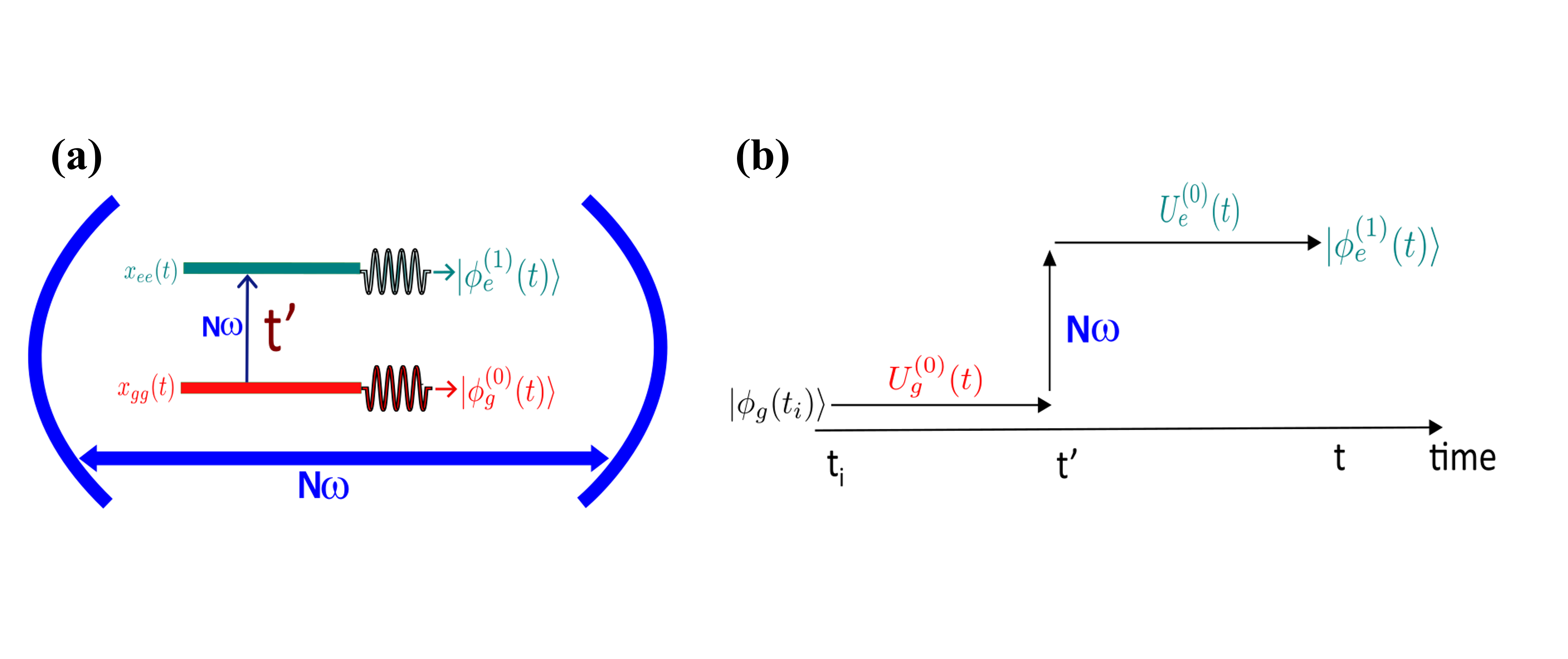}
\caption{
Generation of entangled harmonics correlated to the dressed excited state of the material system. (a) Resonant excitation, which can occur at any time $t'$, alters the nonlinear material response and thus alters the generated product state of harmonics
correlated to a specific excitation time $t'$. The overall state $|\phi_e(t)\rangle$ is an integral, i.~e. a sum, over all excitation times $t'$, and is therefore no longer a product state.
(b) Cartoon diagram describing the generation process, with the states of the harmonic fields evolving under the propagator $U_g^{(0)}$ from $t_i$ until $t'$ and $U_e^{(0)}$ from $t'$ to $t$.
Since excitation can happen at any time, quantum mechanics demands summation over $t'$, resulting in the sum over the product states, i.e. in an entangled state.
} 
\label{Figure2}
\end{figure*}

 Eq.(\ref{eq:SolutionFirstOrderE}) makes two 
important approximations. First, it assumes that the dipoles $\mathbf{x}_{gg}(t)$, $\mathbf{x}_{ee}(t)$ oscillate predominantly at harmonic 
frequencies. This implies that $\ket{\psi_g(t)}$ and $\ket{\psi_e(t)}$ are two Floquet states, consistent with adiabatic turn-on of the driving field with a broad envelope $f_1(t)$, see Eq.(\ref{eq:FieldOperatorSP_1}). Second, it uses perturbative approach, which implies that the probability of 
excitation by the harmonic photon is small compared to unity.
At the same time, sufficient number of harmonic photons
should be generated by the material system to ensure
that accounting for its excitation is meaningful.
In a gas with $N_a$ atoms, the rate of harmonic generation
scales as $N_a^2$, while the rate of harmonic absorption scales with $N_a$, ensuring
that sufficient number of harmonic photons can be generated
before the perturbative approximation breaks down.

The physical meaning of Eq.(\ref{eq:SolutionFirstOrderE}) is simple and sketched in Fig.2. From the 
initial time $t_i$ until some instant $t'$ the 
quantum fields at all harmonic frequencies are
driven by the dipole response $\textbf{x}_{gg}(t)$ generated by the material system in 
its dressed ground state (which includes all 
dynamics induced by $\textbf{F}_{\rm cl}(t)$, possibly even the full depletion of the field-free ground state.)
At an instant $t'$ all quantum fields
at all harmonics are "lifted" by the operator 
$\textbf{x}_{eg}(t')\hat{\textbf{F}}_{I}(t')$ to the new 
electronic "surface", where they evolve from $t'$ to $T$ driven by the 
dipole $\textbf{x}_{ee}(t)$ correlated to the dressed excited state.

Given the near-resonant (or resonant) condition for the 
harmonic $n_0\omega$, we shall 
use the resonant (rotating wave) approximation for
the one-photon transition driven by this harmonic.
This means that the full coupling term $ \textbf{x}_{eg}(t')\hat{\textbf{F}}_{I}(t')$ is approximated by its resonant
contribution, 
\begin{eqnarray}
  &&  \textbf{x}_{eg}(t')\hat{\textbf{F}}_{I}(t')\simeq 
    \Tilde{\textbf{x}}_{eg} 
    e^{+i\int_{t_i}^{t'} d\tau \omega_{eg}(\tau)}
    \boldsymbol{\mathcal{F}}_{n_0}
    \left[\hat{a}_{n_0}e^{-in_0\omega t'}
    \right]=
    \nonumber \\
   && =
    \Tilde{\textbf{x}}_{eg} 
    e^{+i\int_{t_i}^{t'} d\tau \Delta_{eg}(\tau)}
   \boldsymbol{\mathcal{F}}_{n_0}\hat{a}_{n_0},
\label{eq:ExcitationOperator1}
\end{eqnarray}
where the detuning of resonance $\Delta_{eg}(t)=
\omega_{eg}(t)-n_0\omega$ incorporates the 
Stark shifts of the classically dressed states, and
$\Tilde{\mathbf{x}}_{eg}$ is the dipole transition matrix element at the resonant frequency.
This approximation assumes that electronic excitation
between the dressed states
is accompanied by annihilating the resonant photon $n_0\omega$.
In principle, $\Tilde{\mathbf{x}}_{eg}$ can
retain a slow temporal dependence on the pulse envelope,
$\Tilde{\mathbf{x}}_{eg}(t)$, which 
incorporates additional distortions of the dressed states (beyond the simple Stark shift) induced
by the driving field $\omega$.  

The integrand in Eq.(\ref{eq:SolutionFirstOrderE}) can now be written in a simple form. Indeed, 
the initial quantum state of the harmonic light is a product of
the vacuum states for all harmonics, the propagator $\hat{U_g}^{(0)}(t',t_i)$
is a product of linear zero-order propagators for each individual harmonic, and 
at $t'$ the quantum state of light remains a product state, just as in the zero order discussed above. 
Moreover, just as in the zero order, for each harmonic $n$ the propagator $\hat{U}^{(0)}_{n,g}(t',t_i)$ simply shifts the vacuum state $|\alpha_n(t_i)=0\rangle$
to the coherent state 
$|\alpha_{n,g}(t')\rangle$ with
\begin{eqnarray}
    \alpha_{n,g}(t')=-i 
    \boldsymbol{\mathcal{F}}_n \int_{t_i}^{t'} d\tau \textbf{x}_{gg,n}(\tau). 
\label{eq:Alpha(t')}
\end{eqnarray}
Next, we apply the transition operator Eq.(\ref{eq:ExcitationOperator1}) in Eq.(\ref{eq:SolutionFirstOrderE}). It leaves all $n\neq n_0$ undisturbed, while for $n=n_0$ it yields $\hat{a}_{n_0}|\alpha_{n_0,g}(t',t_i)\rangle=
\alpha_{n_0,g}(t')|\alpha_{n_0,g}(t')\rangle$.
Finally, the action of the propagator $\hat{U}^{(0)}_{n,e}(t,t')$ on the coherent state
$|\alpha_{n,g}(t')\rangle$ yields another coherent state at the moment $t$,
$\hat{U}^{(0)}_{n,e}(T,t')|\alpha_{n,g}(t')\rangle=|\alpha_{n,eg}(T,t')\rangle$, where
\begin{eqnarray}
\label{eq:Alpha(t,t')}
    && \alpha_{n,eg}(T,t')= \\ 
    && =-i 
    \boldsymbol{\mathcal{F}}_n 
    \left[
    \int_{t_i}^{t'} d\tau \textbf{x}_{gg,n}(\tau) +  
    \int_{t'}^{T} d\tau \textbf{x}_{ee,n}(\tau)
    \right].
\nonumber
\end{eqnarray}
This yields the final result in the first-order perturbation theory with respect to the quantum 
feedback between the generated light and the material system,
\begin{eqnarray}
\label{eq:SolutionFirstOrderE_RWA_2}
   && |\phi_{e}^{(1)}(T)\rangle=-i\times
    \\
   && \times \int^T_{t_i} dt' \Tilde{\textbf{x}}_{eg} 
    e^{+i\int_{t_i}^{t'} d\tau \Delta_{eg}(\tau)}
    \boldsymbol{\mathcal{F}}_{n_0}\alpha_{n_0,g}(t')
   \prod_{n>1}|\alpha_{n,eg}(T,t')\rangle.
   \nonumber
\end{eqnarray}
The product $\mathcal{F}_{n_0}\alpha_{n_0,g}(t')$
is just the strength of the classical resonant harmonic field at the moment of the 
transition $t'$. 
The matrix element $\textbf{F}_{n_0}(t')\Tilde{\textbf{x}}_{eg}$  is
thus the same matrix element as would have appeared when 
treating the resonant harmonic field as a classical field.

In general, Eq.(\ref{eq:SolutionFirstOrderE_RWA_2}) 
describes a quantum state of light. Indeed, the integral is nothing but a sum of coherent states $\prod_{n>1}|\alpha_{n}(t,t')\rangle$, with  $t'$ a continuous summation index. 
A sum of coherent states of light is no longer
a coherent state. In fact, a sum of coherent states 
can be used to represent highly complex quantum states
of light -- 
as long as one can control the coefficients in such sum.
In the next section we discuss several specific scenarios that realize different entangled states, demonstrating routes to control these coefficients and thus the resulting outcome. 

Note that all generated states emerge from coherent states characterized by their indices $\alpha_n$ 
Eq.(\ref{eq:Alpha(t')}). Thus, in what follows, 
we will characterize the generated states using the average photon numbers 
\begin{eqnarray}
    N_{n,g}(T)=|\boldsymbol{\mathcal{F}}_n \int_{t_i}^{T} d\tau \textbf{x}_{gg,n}(\tau)|^2
\label{eq:N_n(T)}
\end{eqnarray}
that the coherent states at frequencies $\omega_n=n\omega$ would have acquired at the end of the laser pulse $T$, had the material system stayed in the dressed ground state throughout the whole laser pulse.

In particular, we will  use $N_{3,g}\equiv N_{3,g}(T)$
for the third harmonic correlated
to the dressed ground state of the material system as the overall reference number.
We fix this number when presenting quantitative plots of the generated
quantum fields.
For  other harmonics, the key numbers are  $N_{n,g}(T)
\equiv N_{n,g}$, which
are controlled by the parameters of the laser pulse and, if a cavity
is used, by the parameters of the cavity. For example, if $n=5$ is the harmonic resonant with the 
cavity mode and the transition frequency in the material system, the driving laser intensity, and 
the quality of the cavity can allow one to vary 
$N_{5,g}/N_{3,g}$ in a broad range.  In the specific case of Na atom considered in Section 4D, we obtain all relevant quantities from
the numerical solution of the time-dependent Schrödinger equation; fixing 
$N_{3,g}$ then automatically defines $N_{5,g}$, etc. and allows us to
present quantitative results, see Section 4D.

\section{Results and Discussion}

\subsection{Estimates}

We shall see below that strong quantum effects in the generated harmonic light emerge from 
the semi-classical harmonic generation regime 
for $N_{n_0}(T)\sim 10-10^2$ photons in the resonant harmonic
$n_0$ generated 
in the coherent state.
To get a feeling for the typical parameters required
to reach $N_{n_0}(T)\sim 10-10^2$, 
consider a vapor of Na atoms. We have chosen 
Na because of the very strong
transition between its ground 3s and the first excited 3p 
state, with the transition wavelength near $\lambda_0\simeq 590$ nm. In a micro-cavity with a volume $V\sim \lambda_0^3$ and a quality Q-factor $Q\sim 10$, 
a coherent state with $N\sim 10^2$ photons 
at $\lambda_0$ will yield a resonant field with
intensity $I\sim 3\times 10^7$W/cm$^2$. 
We shall see below that this intensity of the resonant field 
is sufficient to
dramatically alter harmonic generation
driven by the fundamental field with a modest 
intensity of $I\sim 3\times 10^{11}$W/cm$^2$
carried at $\lambda=1770\simeq 3\times 590 $ nm,
so that its third harmonic is one-photon resonant with 
the 3s-3p transition.

The TDSE simulations 
for a single Na atom presented below show that even a single Na atom in its dressed ground state will generate $N_{3,g}\sim 10$ photons in about 10 psec,
for a $V\sim \lambda_0^3$ cavity with $Q\sim 10$.
This time can be shortened to $\sim 100$ fsec by placing
$N_a\sim 10$ atoms in a cavity, even without any quantum
correlation between them. Even without 
cavity enhancement, i.e. for $Q=1$, $N_a\sim 10^2$ Na atoms within $\lambda_0^3$ volume would be sufficient to generate
$N_{3,g}\sim 10^{2}$ photons in the resonant mode within $\sim 10^2$ fsec.  This number of atoms 
corresponds to  $10^{14-15}$ cm$^{-3}$ number density, 
rather low by the standards of 
conventional harmonic generation experiments.
Thus, already within the standard semiclassical harmonic 
generation regime, sodium vapor at relatively 
modest conditions will generate enough resonant 
harmonic photons to enable quantum effects. Our TDSE simulations for Na presented below are performed for 
40 fsec pulses and assume sufficient vapor density to generate 10 resonant photons in the semiclassical regime,
during this time. The experiment  \cite{jackson82}
shows that generation of sufficiently intense
resonant third harmonic in standard regime is feasible.
In the Dicke superradiance regime, the outcome might 
even be more interesting. However, analysis of the opportunities 
to realize the Dicke regime is outside the scope of this paper.

\subsection{Quantum light generation via Freeman resonances}

To appreciate the entangled nature of the harmonics
Eq.(\ref{eq:SolutionFirstOrderE_RWA_2})
correlated to the excited state, as well 
as the possibility to control the state of the 
generated light, we begin by considering
the so-called Freeman resonances \cite{freeman1987above, gibson1992verification}. These refer 
to the situation where the conventional Stark shift proportional to the intensity of the classical IR driver brings the states in and out of resonance, as shown in Fig.3(a). The integral is now accumulated during the temporal windows near the resonant times $t_k$ and can be computed using the standard stationary phase method, yielding
at the end of the laser pulse $T$
\begin{eqnarray}
&&    |\phi_{e}^{(1)}(T)\rangle=\sum_{k} C_k e^{+i\int_{t_i}^{t_k} d\tau \Delta_{eg}(\tau)} 
   \prod_{n>1}|\alpha_{n,eg}(T,t_k)\rangle,
   \nonumber \\
&&   C_k=\sqrt{\frac{-2\pi i}{\dot\Delta_{eg}(t_k)}}\Tilde{\textbf{x}}_{eg}  
\boldsymbol{\mathcal{F}}_{n_0}\alpha_{n_0,g}(t_k) \ .
\label{eq:SolutionFirstOrderE_RWA_3}
\end{eqnarray}
Here $\dot\Delta_{eg}(t_k)$ is the time-derivative of the resonance detuning at the 
moment of resonance $t_k$. The coefficients $C_k$ are nothing but the amplitudes of one-photon excitations by the resonant harmonic $n_0$ during the resonance windows $t_k$. 
As discussed above, they are the same as for
the classical harmonic field with the amplitude $F_{n_0}(t_k)=\mathcal{F}_{n_0}\alpha_{n_0,g}(t_k)$
at the moment $t_k$. 

The phase factors
$e^{+i\int_{t_i}^{t_k} d\tau \Delta_{eg}(\tau)}$ control the relative phases between the terms in the sum. Overall, 
the amplitudes $C_k$ describe the non-adiabatic Landau-Zener-Dykhne transitions between the dressed states
driven through the resonance, with the phase factors  $e^{+i\int_{t_i}^{t_k} d\tau \Delta_{eg}(\tau)}$
responsible for the Stueckelberg oscillations (named so after  
baron Ernst C. G. Stueckelberg von Breidenbach) in the final generated state, 
as the coherent states are not orthogonal.
We stress again that both the number of the terms in the sum and the relative phases between them are controlled by shaping the intensity envelope of the fundamental laser pulse.

\begin{figure*}[htp]
\centering
\includegraphics[width = 1.0\linewidth]{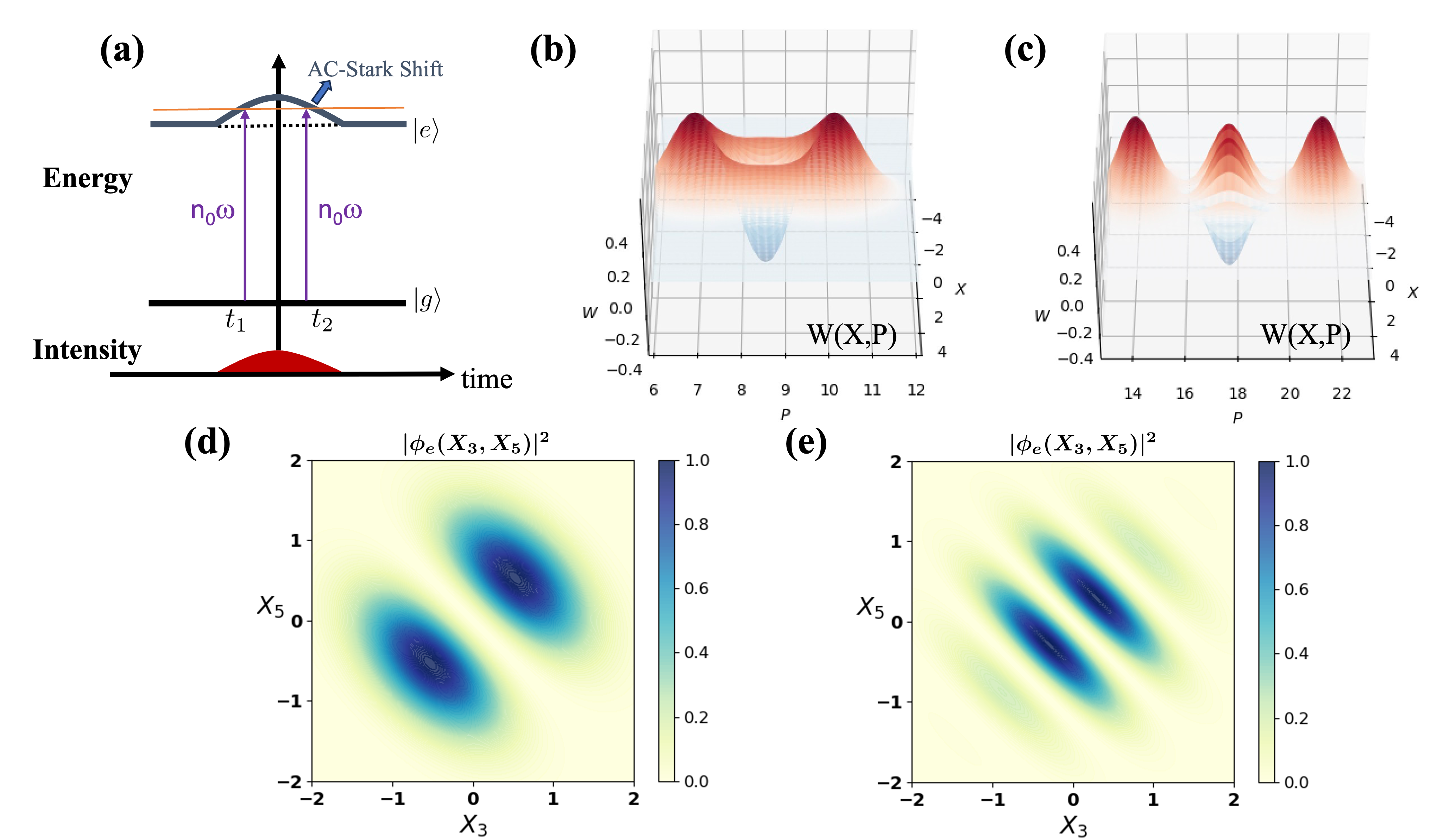}
\caption{
Control of resonances and of the generated quantum states of light. (a) Stark shift follows the intensity envelope of the driving laser field and brings the ground and excited states into resonance twice during the pulse. 
(b,c) Wigner function for one of the harmonics, here $n=3$, upon artificially removing its entanglement with others. For illustration, we set resonances at $t_1$ and $t_2$ to occur when the average number of the harmonic photons $N_3(t)$ generated by the laser-dressed ground state are $N_3(t_1)=25$, $N_3(t_2)=50$ (b) and $N_3(t_1)=100$, $N_3(t_2)=225$ (c).  (d,e) Intensity of the $X$-quadratures of the entangled state of two harmonics, here $n=3$ and $n=5$, 
$|\phi_e(X_3,X_5)|^2$. The parameters are set to $N_3(t_1)=N_5(t_1)=25$, $N_5(t_2)=N_3(t_2)=50$ in (d) and $N_3(t_1)=N_5(t_1)=100$, $N_5(t_2)=N_3(t_2)=225$ in (e).} 
\label{Figure3}
\end{figure*}

Examples of quantum states of light that can be created this way are shown in Fig.3, for two resonant windows and hence two terms in the sum. 
For simplicity of visual presentation, we assume that
only two harmonics, the third (n=3) and the fifth (n=5), are efficiently generated, so that
the entangled wavefunction contains only two terms in each product state, 
\begin{eqnarray}
    \nonumber
    \ket{\phi_{e}^{(1)}(T)}= && C_1 \ket{\alpha_{3,eg}(T,t_1)} \ket{\alpha_{5,eg}(T,t_1)}+
    \nonumber \\
    &&+C_2 e^{i\Phi}\ket{\alpha_{3,eg}(T,t_2)} \ket{\alpha_{5,eg}(T,t_2)},
     \nonumber \\
     \Phi= &&\int_{t_1}^{t_2} d\tau \Delta_{eg}(\tau).
     \label{eq:SolutionFirstOrderE_RWA_4}
\end{eqnarray}
This constraint is
only used for simplicity of visualizing the 
overall quantum state of light. Note that $\alpha_n$, $C_{1,2}$ and $\Phi$ are fully controlled by the parameters of the driving laser pulse and the medium.

\begin{figure*}[htp]
\includegraphics[width = 1\linewidth]{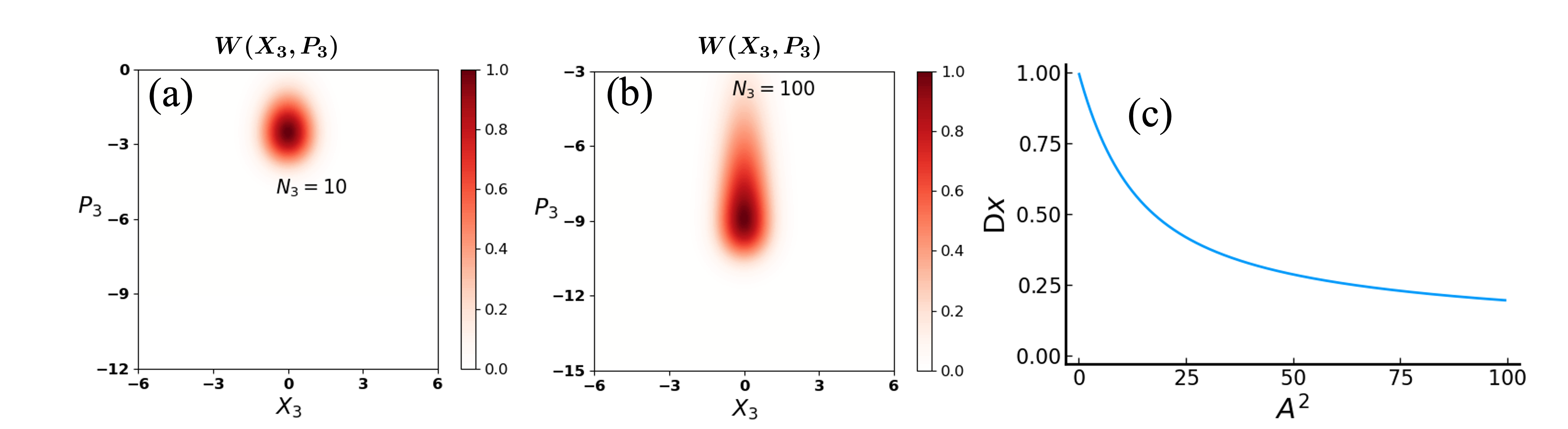}
\caption{
Examples of the single harmonic state correlated to the excited state in case of resonance at all times.  (a,b) Wigner functions for one of the harmonics, $n=3$, upon removing its entanglement with others, for the average number of photons $N_3(T)=10$ (a) and $N_3(T)=100$ (b). (c) The dispersion $\text{D}X$ given by the wavefunction Eq.(\ref{eq:Chi(X)}) for
a single harmonic, here $n=3$. $DX$ is normalized to that of the vacuum state. For $x_{ee,n}=0$, $A^2$ is equal to the average number of photons $N_{3,g}$ defined in Eq.~(\ref{eq:N_n(T)}). The dispersion of the $x$ quadrature is sub-classical for any nonzero value of $A^2$ ($N_{3,g}$).
} 
\label{Figure4}
\end{figure*}

To give an impression of the underlying structure of a single frequency mode, say $n=3$, we first artificially set $\alpha_{5,eg}(T,t_1)=\alpha_{5,eg}(T,t_2)$, which allows us to disentangle
the third harmonic from the fifth and plot its Wigner function.
 The resulting state is, clearly, a Schrödinger cat-like: $C_1\ket{\alpha_{3,eg}(T,t_1)}+C_2e^{+i\Phi}\ket{\alpha_{3,eg}(T,t_2)}$. 

To visualize this expression, 
we need to specify the coherent states 
$\ket{\alpha_{3,eg}(T,t_1)}$ and $\ket{\alpha_{3,eg}(T,t_2)}$, i.e.
the photon numbers $N_{3}(t_1), N_{3}(t_2)$ 
associated with them, as well as the relative phase.
Fig.3 shows the Wigner functions  
for the cases of $\Phi=\pi$ and $N_{3}(t_1)=25$,  $N_{3}(t_2)=50$ in panel (b) and $N_{3}(t_1)=100$,  $N_{3}(t_2)=225$
in panel (c). 
For a symmetric driving pulse, the 
one-photon resonant excitation amplitudes $C_1$ and $C_2$ 
at the front and rear ends of the driving pulse 
are similar; in Fig.3 we set them to be equal.
Note that both the relative phase $\Phi$ and the photon numbers are fully controlled by the time delay between the resonances and the Stark shifts induced by the driving laser field.  The characteristic non-classical features of the   
cat-like state are apparent in Figs. 3(b,c). 
We note
once more the importance of the relative phase factor $e^{i\Phi}$ in the
cat-like state -- it is fundamentally different from the classical
phase factor $e^{i\mathrm{arg}(\alpha)}$ of every particular coherent state
$\ket{\alpha}$. The relative phase $\Phi$ defines the interference fridges seen in Fig. 3;
In our case, $\Phi$ is controlled by the pulse duration and shape.

We next move to the example where both the third and the fifth harmonics are generated with similar
efficiency. This could be achieved, for example, by tuning the cavity resonance close to the $n=5$ harmonic
and staying within the perturbative regime with respect to the driving laser field, so that higher order harmonics
are negligible. The 2D plot of the resulting
state as a function of the x-quadratures $X_3$ and $X_5$ of both harmonics is shown in Fig.3(d), for
$N_{3,g}(t_1)=N_{5,g}(t_1)=25$ and $N_{3,g}(t_2)=N_{5,g}(t_2)=50$, with equal amplitudes $|C_1|=|C_2|$ and the relative
phase $\Phi=\pi$ between the two terms. Panel (e)
shows the same for $N_{3,g}(t_1)=N_{5,g}(t_1)=100$ and $N_{3,g}(t_2)=N_{5,g}(t_2)=225$.

\subsection{Exact resonance: analytical results}
 
Consider a driving pulse with constant peak intensity
turned on at $t_i=0$ and off at $T$, realizing exact resonance between the Stark-shifted Floquet 
states $|\psi_{e,g}\rangle$ at all times from $t_i=0$ to T.
The  integral Eq.(\ref{eq:SolutionFirstOrderE_RWA_2}) becomes
\begin{eqnarray}
\label{eq:SolutionFirstOrderE_RWA_Resonant}
   && |\phi_{e}^{(1)}(T)\rangle=-i\times
    \\
   && \times \int^T_{t_i} dt' \textbf{x}_{eg,n_0} 
    \boldsymbol{\mathcal{F}}_{n_0}\alpha_{n_0,g}(t')
   \prod_{n>1}|\alpha_{n,eg}(T,t')\rangle.
   \nonumber
\end{eqnarray}
where $\textbf{x}_{eg,n_0}$ is the Fourier component of
the transition matrix element $\textbf{x}_{eg}(t)$ at the resonant frequency.

To compute
the integral, we need the Fourier components
$x_{gg,n}$, $x_{ee,n}$ and $x_{eg,n}$. From now on, we assume them to be 
time-independent, obtained by Fourier transforming the 
nonlinear response across the whole pulse. 
In general, $x_{gg,n}$ and $x_{ee,n}$  are obtained
from numerical simulations of the time-dependent 
Schrödinger equation (TDSE), for the material system
starting in the 
ground and excited states. In the next subsection, we will 
use such simulations for the specific 
case of a sodium atom. Here, we  shall
consider in a general form the case of a 
low-frequency driver
at modest intensity and for negligible ionization, with 
$x_{gg}(t)$ and $x_{ee}(t)$
in phase (or exactly out of phase) with the driving 
field. The Fourier components, $x_{gg,n}$ and $x_{ee,n}$, 
are then real and the integral can be easily computed using quadrature representation of the coherent states. The 
result is
\begin{eqnarray}
\label{eq:SolutionFirstOrderE_RWA_Resonant3}
    \ket{\phi_{e}^{(1)}(T)} =
   -{x}_{eg,n_0} x_{gg,n_0}
    {\mathcal{F}}_{n_0}^2 \frac{T^2}{2}
    \ket{\eta_e(T)}.
\end{eqnarray}
In the \textbf{X}-quadrature representation,
$\textbf{X}=X_3, X_5, ...$, the wavefunction $\ket{\eta_e(T)}$
can be written as
\begin{eqnarray}
\label{eq:Chi(X)}
 \eta_e(\textbf{X},T)=\xi_e(\textbf{X},T)\prod_{n>1}
 \varphi_{\alpha,n}(X_n,T). 
\end{eqnarray}
The function $\varphi_{\alpha,n}(X_n,T)$ is the coherent state of $n$-th harmonic, describing  the shift of the initial vacuum states that would have been generated by the excited state alone, 
$|\varphi_{\alpha,n}(T)\rangle\equiv \ket{\alpha_{n,e}(T)}$, where
$\alpha_{n,e}(T)=-i{\mathcal{F}}_{n}x_{ee,n}T$. The term 
$\xi_e(\textbf{X},T)$ is more interesting, as it cannot be written as a product state for individual harmonics:
\begin{eqnarray}
\label{eq:Xi(X)}
&& \xi_e(X_3,X_5,...,T)=e^{-i\textbf{A}\cdot \textbf{X}}\frac{1+i\textbf{A}\cdot \textbf{X}
-e^{i\textbf{A}\cdot \textbf{X}}}{(\textbf{A}\cdot \textbf{X})^2},
\\
&& \textbf{A}\cdot \textbf{X}=\sum_{n>1}A_n X_n= \sqrt{2}\sum_{n>1}X_n\mathcal{F}_{n}(x_{gg,n}-x_{ee,n})T.
\nonumber 
\end{eqnarray}
Once $x_{gg,n}$, $x_{ee,n}$ are known, the components 
$A_n=\sqrt{2}\mathcal{F}_{n}(x_{gg,n}-x_{ee,n})T$ of the
vector $\textbf{A}$ can be calculated and the resulting quantum states can be  
fully quantified. 

Note that $\xi_e(\textbf{X},T)$ is equal to unity if $\textbf{A}=0$. The limit $\textbf{A}=0$ corresponds to the case when excitation of the material system makes no difference to harmonic generation ($x_{ee,n}=x_{gg,n}$); the harmonics remain
in the product of coherent states. However, if the 
nonlinear response of the ground and excited states
is different, nontrivial entangled quantum states are generated.

\begin{figure*}[htp]
\includegraphics[width = 1\linewidth]{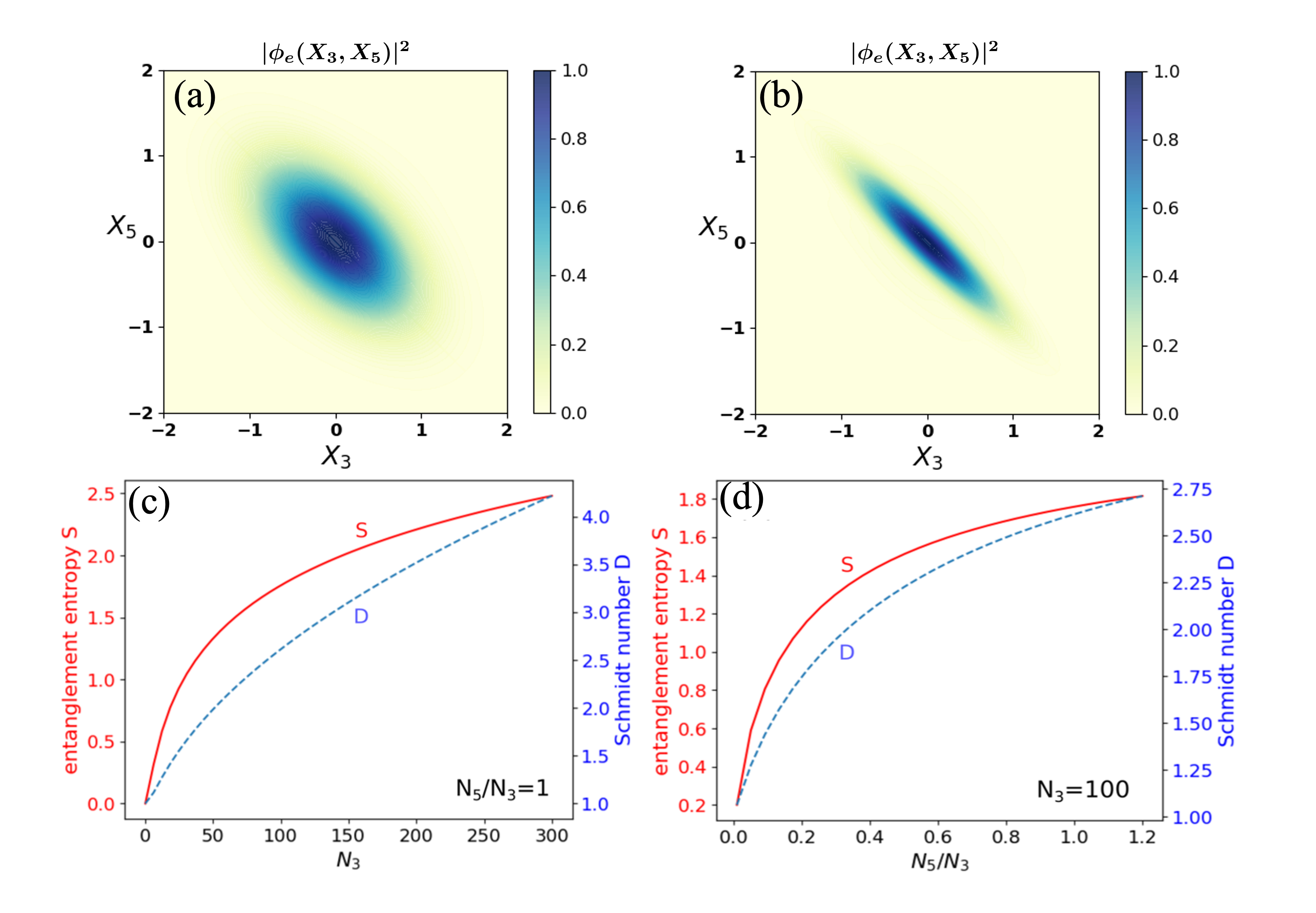}
\caption{ The entangled state between 3rd and 5th harmonic modes. 
(a,b) Intensities of the entangled wavefunctions in X-quadratures for $N_3(T)=N_5(T)=10$ (a) and $N_3(T)=N_5(T)=100$ (b). 
(c,d) The entanglement entropy $S$ and the Schmidt number $D$ as functions of the average number of photons $N_3$ and $N_5$, $N_3$ is changed, keeping constant the ratio $N_5/N_3$ (c) and the ratio $N_5/N_3$ is modified, keeping $N_3$ constant (d). 
} 
\label{Figure5}
\end{figure*}

Examples of such states  are shown in Fig.\ref{Figure4} and 
Fig.\ref{Figure5}.
For illustrative purposes, here we assume the nonlinear response of the excited state to be negligible, $x_{ee,n}=0$.
This is reasonable for a resonance with a Rydberg state in a modest field regime, before
the onset of recollision-driven harmonics.
Indeed, the optical response of a Rydberg state is nearly
identical to that of the free electron, with negligible nonlinearity before the onset of recollision harmonics.
The case of 3s, 3p states of Na, where the nonlinear response
in the excited 3p state is stronger than in the ground 3s state, is illustrated in Fig.6.

To give an impression of the underlying structure of a single frequency mode, say $n=3$, we first suppress the 
generation of all harmonics except for the third, and plot its Wigner function of the third harmonic
for $N_{3,g}=|\alpha_{3,g}(T)|^2=10$ and 
$N_{3,g}=|\alpha_{3,g}(T)|^2=100$, see Fig.4(a,b). 
We see that the harmonic becomes progressively more squeezed as the number of photons increases. 
The degree of squeezing is quantified in
Fig.4(c), where we plot the $X_3$-quadrature dispersion of the
wavefunction $\eta_e(X_3,T)$ as a function of $A^2$. The dispersion is normalized to that of the vacuum state.

Fig.4(c) shows that the $X_3$-quadrature dispersion decreases significantly below that for the vacuum state, with squeezing 
increasing with increasing $A$, which measures
the difference in the nonlinear response of the 
two dressed states. While Fig.4(c) shows
a specific case of a single mode, the result is general:
for multiple modes (multiple harmonics), multi-mode
squeezing develops along the vector $\mathbf{A}$.
The non-Gaussian character of the resulting state signifies it as the inherently quantum resource 
\cite{hudson74, albarelli18}.

While the growing squeezing is already very interesting, the overall quantum state Eq.(\ref{eq:Chi(X)}) is richer.
Fig.\ref{Figure5} shows the case when two harmonics, $n=3$ and $n=5$, are efficiently generated. The developing quantum correlation is already apparent from the state intensity $|\phi_e(X_3,X_5)|^2$ shown as a function of the 
$X$-quadratures for both harmonics, which becomes both 
squeezed and rotated -- a tell-tale sign 
of a correlated state. 
To characterize entanglement, we use 
the Schmidt number and/or the entanglement entropy.
Both are obtained from the Schmidt decomposition of $\phi_e(X_3,X_5)$
as a sum over the separable states \cite{law00,grobe94}:
$\phi_e(X_3,X_5)=\sum_i\lambda_i \phi_{e,i}(X_3)\phi_{e,i}(X_5)$,
where $i=1,2,3,\ldots$, $\phi_{i}(X_3)$, $\phi_{i}(X_5)$
represents a set of orthonormal single-harmonic wavefunctions, and $\lambda_i>0$ are the corresponding eigenvalues of the Schmidt decomposition, normalized so that $\sum_i \lambda_i^2=1$. Entanglement entropy $S$ is defined then as $S=-\sum_i \lambda_i^2\log_2\lambda_i^2$, whereas the Schmidt number $D$ is defined \cite{grobe94} as $D=1/{\sum{\lambda_i^4}}$. If the state is separable, we have $S=0$ and $D=1$, whereas for non-separable states, $S>0$, $D>1$. Both quantities show how entangled the states are. 
For the generated states, these measures are shown in 
Figure \ref{Figure5}, clearly 
demonstrating strong photon-number entanglement developed between harmonics.
Importantly, the degree of entanglement and non-classicality of harmonics grows with increasing the average number of photons
per harmonic, a very exciting trend.

\begin{figure*}[htp]
\includegraphics[width = 1\linewidth]{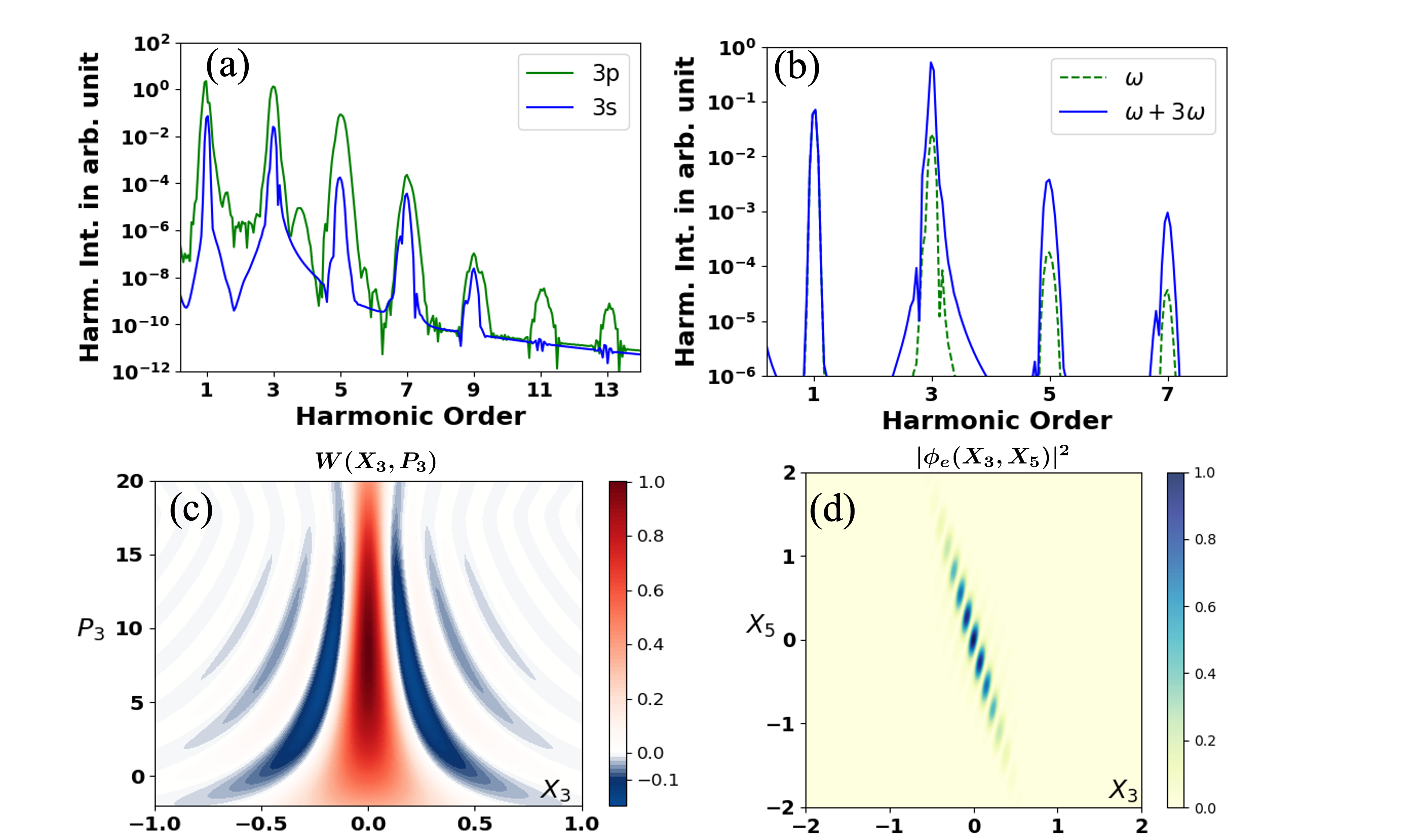}
\caption{ Harmonic emission in Na from the solution of the time-dependent Schroedinger equation,
TDSE. (a) Harmonic intensities generated when starting in the ground 3s (blue curve) and the first excited 3p state (orange curve) of Na, for $\lambda=1770$ nm 40 fsec FWHM pulse with a Gaussian envelope at intensity of $3\times10^{11}$ W/cm$^2$. (b) Harmonic emission in Na starting in the ground 3s for a single-color 1770nm pulse (green) with the same parameters as in (a), and for a two-color pulse (blue), with the second color at 3rd harmonic with intensity of $I=3\times 10^7$W/cm$^2$, $10^{-4}$ of the fundamental, with zero relative two-color phase. (c) Wigner function for the 3rd harmonic correlated to the 3p state, 
for the average number of photons $N_{3,g}(T)=10$
generated in the ground state, which corresponds to $N_{3,e}(T)=794$ in the excited state. (d) Intensity of the entangled wavefunctions
for harmonics 3, 5 correlated to the 3p state in X-quadratures for $N_{3,g}(T)=10$, which corresponds to 
$N_{5,g}(T)=0.1$, $N_{3,e}(T)\simeq 794$, and $N_{5,e}(T)\simeq 65$.}    
\label{Figure6}  
\end{figure*}

\subsection{Application to Na atom}

We now consider sodium atom, where there is a strong transition from the ground 3s to the first excited 3p state near $\lambda$=590 nm. 
The calculations were done using time-dependent Schrödinger 
equation in the single active electron approximation, including
ionization; the technical details are described in the Appendix. 
We centered the driving laser pulse at $\lambda=1770$ nm,
tuned to the three-photon resonance with the transition. We kept
the intensity of the driving field modest to minimize
ionization and limit harmonic generation to
the regime well below the onset of the re-collision and 
Brunel-type \cite{brunel1990harmonic} harmonics.
The results for the intensity of $3\times 10^{11}$ W/cm$^2$ and 40 fs (full width at half-maximum, FWHM, 
truncated Gaussian\cite{Patchkovskii2016} envelope) pulse duration are shown in Fig.6(a,b) and are typical for this parameter range.  

First, we ensure that we can produce the resonant harmonic H3 efficiently, starting with Na in the ground 3s state, see the blue curve in Fig.6(a). The third harmonic response is nearly as strong as the linear response, ensuring that an intense third harmonic can indeed be produced in typical experimental conditions. 

Second, we have seen above that the key condition for creating non-classical states is to ensure different nonlinear responses correlated to the ground and excited states. To check this, we calculated harmonics starting from the 3p excited state, see the green curve in Fig.6(a). We see that in Na the intensities of the nonlinear optical response correlated to the excited 3p state are almost two orders of magnitude higher for harmonic 3, and even higher for harmonic 5. The second condition is thus met.

Third, we check how much third harmonic intensity we need to provide enough resonant interaction and  noticeably alter the harmonic emission. Fig.6(b) compares the results for the fundamental driver at 1770 nm with the intensity of $I=3\times 10^{11}$ W/cm$^2$ (green curve) to the case of 
a two-color field containing both the fundamental and the third harmonic (blue curve), which is resonant with
the 3s-3p transition. In both cases, the calculations start
in the 3s (ground) state.
Even when the third harmonic intensity is as low as $I=3\times10^7$ W/cm$^2$, 10$^{-4}$ of the fundamental, the harmonic emission is altered dramatically, with harmonics H3, H5, and H7 increased by about an order of magnitude.

Thus, our calculations show that generating third harmonic centered at 590 nm in Na with an intensity $\sim 10^7$ W/cm$^2$ is sufficient to have major effect on all harmonics. 
Accumulating third harmonic with such intensity is well within reach in a standard experiment, even without
a cavity.  Using a typical 10$^2$ micron focal radius of the fundamental and thus having the third harmonic beam of about 10$^2$ micron diameter, we need to generate about $10^7$ photons in this beam. Given that $10^6$ photons are routinely obtained for high harmonics deep in the XUV range, the required conditions are fully realistic.  This conclusion agrees with the estimates given at the beginning of this Section. In fact, the dramatic effect of resonant third harmonic 
generated in Xe gas on photo-ionization was observed already in 1982 \cite{jackson82}, where a coherent
superposition of the dressed ground and excited states
was created by the generated third harmonic 
to suppress photo-ionization.

Fig.6(c) shows the Wigner function for the quantum state of the third harmonic correlated to the excited dressed state, for $N_{3,g}=10$.
The rest of the parameters are extracted from the TDSE simulation for Na. In particular, at this intensity 
the optical response 
of the excited 3p state is out of phase with that of the 
ground 3s state, with $x_{ee,3}/x_{gg,3}=-8.91$. 
For simplicity, we assumed constant driving pulse intensity
and used Eq.(\ref{eq:Chi(X)}). To plot
the Wigner function for the third harmonic, we have
artificially suppressed the nonlinear response at higher harmonics.
In contrast to the cases shown in  Figs.~3,4,  
here $|\textbf{x}_{gg,n}|\ll|\textbf{x}_{ee,n}|$  
and the Wigner function develops small negative values
(shown as blue areas in Fig.6(c)), demonstrating 
highly nonclassical nature of the 3rd harmonic.

Once we include the response at the fifth harmonic, we can plot the corresponding wavefunction
in the quadratures $X_3,X_5$, for the same laser 
parameters  (Fig. 6(d)). 
Given that the fifth harmonic response in the 
ground state is two orders of magnitude weaker than
for the  third  harmonic (see Fig.6(a)), $N_{3,g}=10$
corresponds to $N_{5,g}(T)=0.1$. However, the excited
state response is much  stronger,  
$x_{ee,5}/x_{gg,5}=-25.53$,
yielding $N_{5,e}(T)\simeq 65$. 
As in the case of Fig.5(a,b), this state is clearly entangled, demonstrating, however, in addition, interference fringes, indicating mode-to-mode quantum beatings.

\section{Conclusions and Outlook}

Our work leads to several important conclusions. We find that 
non-trivial quantum states
of harmonic light are generated as soon as
quantum correlations develop between 
the generated light and the generating material 
system. A sufficient condition to developing these correlations is
rather general and quite straightforward:
(i) the system should undergo non-adiabatic excitations between 
laser dressed states and (ii) the different
laser-dressed states should
exhibit different non-linear responses. If this is the case,
excitations of the material system into different
dressed states would generate not only an entangled
light-matter wavefunction, but also an entanglement
between the different harmonics correlated
to the excited state of the material system. 

The quantum states of harmonics emerge
whenever the material excitation occurs at different 
times throughout the laser pulse. Its origin is
the interference of multiple quantum pathways leading to
the same final state of the material system, each 
pathway correlated to different coherent states of
light it generates.

The specific mechanism of resonant excitations between 
different laser-dressed states can be quite general and 
does not need to be limited to the absorption of a generated
harmonic. Resonant non-adiabatic transitions in 
a strongly driven system, with large time-dependent Stark shifts, are virtually inevitable, making the 
process considered here rather typical.

High harmonics correlated to real excitations of matter are 
often ignored due to their poor phase matching properties. Indeed, 
excitation of loosely bound Rydberg states implies
the so-called long trajectories in recollision-driven \cite{schafer1993above,Corkum1994} high harmonics. The contribution of these long trajectories to the macroscopic response is generally weak due to the large intensity-dependent
phase associated with these trajectories \cite{lewenstein1995phase,gaarde1999spatiotemporal}. However, recent 
experiments and theory \cite{mayer2022role} show that this need not be the case, and that harmonics generated via excited
states can even dominate the far-field spectrum. Nevertheless, to improve 
phase matching it might be beneficial to
look at intermediate field regimes before the onset of highly non-perturbative recollision-type
harmonic generation mechanisms, reducing the complexities of phase-matching large intensity-dependent harmonic phases
accumulated during electron motion in the continuum. 

The presence of the cavity is highly beneficial but not strictly necessary: the essential aspect is the significant
excitation probability. In this context, solids, and
  especially two-dimensional 
solid state materials, where excitons provide strong contribution to the optical response, appear an interesting candidate for generating non-classical light, see
also recent work \cite{theidel2024evidence}, especially taking into
account that solids are becoming a standard target
for intense light-matter interactions.  If the cavity
is used, the optimal cavity
lifetime should be on the scale of the pulse duration:  on the one hand, shorter pulses do not interact efficiently with the cavity; on
the other hand, the cavity should tolerate relatively large
harmonic linewidths.

An interesting opportunity is offered by alkali atoms, such as 
sodium discussed here.  The idea of using sodium 
vapors, with or without a cavity, may look like a suggestion to turn the clock back a few decades, returning to a well-studied system. However, we suggest to add a new twist: instead of driving the atomic medium with a resonant field, here we suggest to drive it with a low-frequency field, tuning the laser
frequency into a multiphoton resonance with the transition between the ground and the first
excited states of an alkali atom.

Control over the strengths and times of resonant excitations
becomes a way of controlling the generated entanglement.
We have illustrated this point by using the so-called Freeman resonances \cite{freeman1987above}, which are both ubiquitous and unavoidable during intense light-matter interaction. These resonances are induced by shifting an excited state into a resonance with the ground state due to the ponderomotive shift $U_p=F^2/4\omega^2$
($F$ is the field strength), associated
with the wiggling energy of a loosely
bound electron driven by the incident laser field. 
The resonance is multi-photon 
for the driving field and  one-photon for 
one of the
generated harmonics. Crucially, already a single resonant harmonic
can trigger generation of non-trivial quantum
states for all other harmonics: all it takes is for the resonant
harmonic to invest its photons in material excitation, benefiting all others.

Even in the simplest case of a continuous resonance occurring throughout the whole
laser pulse, which corresponds to a perturbative regime where the ponderomotive Stark shift is negligible, all
harmonics correlated to the system left in the excited state are generated in nontrivial quantum states.
What is more, in contrast to normal expectations, the quantum properties of the generated harmonics become stronger as the number of photons in each harmonic grows.  In particular,
our results show that harmonics become strongly
squeezed, with squeezing increasing with the number of generated photons.

Since Stark shifts, which control resonances in low-frequency laser fields, depend on the laser intensity, shaping the intensity envelope of the driving field shapes the quantum nature of the generated harmonic fields. Whereas here we have only considered the case of two resonances (at the front and at the back of the pulse), the overall number of resonances can be controlled by shaping the laser pulse envelope in time. 
As long as the corresponding modulations are much slower than the driving laser period, the approach we have developed above applies equally well to this case.

How can the nonclassical states of light correlated to the excited dressed state of matter be observed, and does it require post-selection on the 
excited state? 
One option is to generate enough of it, on par with light correlated to the ground state. The example of sodium atoms
considered above shows this to be possible. In sodium, the nonlinear optical
response of the excited state can exceed 
that of the ground state by two to three orders of magnitude, Fig.6(a). Thus, even low degree
of excitation can generate enough quantum light,
on par with the coherent states generated by the dressed ground state. In this, admittedly most favourable case, no post-selection is needed.

But what if the harmonics correlated to the excited state are weak? One way to observe them is to structure matter excitation in space by creating transverse excitation grating.  To this end, 
one could take advantage of a cavity with a periodically structured transverse mode, see Fig.1(c). However, one can 
also envision a sequence of excitation pulses in a non-collinear configuration creating a transient 
grating, similar to the one already used in the experiments \cite{worner2010following} to separate high harmonic emission correlated to the excited state of a dissociating Br$_2$ molecule from the dominant harmonic emission correlated to its ground state. An excitation grating in the near-field ensures that harmonic emission correlated to such excitation forms transverse minima and maxima in the far-field, while the radiation correlated to the ground state is not deflected. 
This separates the 'interesting' light from the standard
semi-classical one.

In our case, a sequence of non-collinear excitation pulses, overlapping
with the pulse driving the harmonic emission, 
will ensure multiple excitation instants and multiple
quantum pathways leading to harmonics correlated to 
the excited state -- the key condition needed for
generating nontrivial quantum states of harmonics.
We leave the detailed analysis of this
idea to a separate paper, presenting here only
a rough sketch. In the near-field, the quantum state of light and matter (see Eq.(\ref{eq:FullWF_1})) is written as
\begin{eqnarray}
\ket{\Psi(t,\rho)}=\ket{\psi_g(t)} \ket{\chi_g(t)} +   \ket{\psi_e(t,\rho)} \ket{\chi_e(t,\rho)}    
\label{eq:NearField}
\end{eqnarray}
where $\rho$ is the transverse coordinate and 
the $\rho$-dependence in
$\psi_e(t,\rho)$ describes the excitation grating created in the material system.
The light wavefunctions  $\ket{\chi_e(t,\rho)}$ include
multiple harmonic modes with different momentum vectors.
For the excitation
grating  with a period $\rho_0$, 
the momentum vector $k_n$ of a harmonic $n$ correlated to this excitation and propagating into the far-field will have transverse 
components $k_n^{(m)}=2\pi m/\rho_0$, with different $k_n^{(m)}$corresponding to different diffraction orders.
This entangles the direction of light propagation with material excitation.  
Selective 
measurement of light in the $m\neq 0$-th diffraction order
will retain full quantum coherence of the harmonics
correlated to the excited state, for arbitrarily large photon numbers. 

Spatial shaping of the generating pulse and/or harmonic modes (e.g. via transverse modes of the resonant cavity) also opens further routes for generating highly nontrivial quantum states of light where the spatial harmonic parameters such as orbital angular momenta are also entangled.

We note that, in general, when considering systems involving a potentially infinite number of vacuum modes, observing only some of those (rather than all) causes partial decoherence of the observed states (see, e.g.~\cite{agudelo2013quasiprobabilities}). For several entangled harmonics, such decoherence will increase with increasing their number, i.e. with increasing 
the intensity of the driving field. The number of excited states in the material system will also typically increase with increasing the laser intensity, entangling multiple harmonics with multiple excited states and with each other. Incomplete measurements 
will then lead to the loss of quantum features with increasing laser intensity, in agreement with the recent experiment \cite{theidel2024evidence}.

Applications of nonclassical, massively entangled states of high harmonics can
be very broad \cite{bhattacharya23}. These include imaging
of material properties, where nonclassical and entangled states of light
provide  quantum advantage \cite{dorfman16}, resulting in, for
instance, enhanced absorption probability and resolution, or selectivity with respect to different quantum pathways in 
wave-mixing spectroscopy. Furthermore,
entangled combs are known to be a resource for quantum information
processing, such as quantum communication and cryptography
\cite{kues19}. Whereas typical combs are produced by electro-optic
modulations of cavities and are limited by GHz bandwidth, the approach
presented here promises to expand it to the PHz frequency range.

This discussion demonstrates truly remarkable 
opportunities arising when combining the fulll toolkit of attosecond physics with the field of quantum optics, with exciting implications in different fields ranging from ultrafast and precision spectroscopy to quantum information science.

\begin{acknowledgments}
    G.B.  acknowledges the support of the Horizon 2020 Framework Programme, award number 899794. M.I. and N. K.
    acknowledge the support of Limati SFB 1477 “Light-matter interaction at interfaces” project, award number 441234705. O.S. acknowledges support of
    the ERC grant 'ULISSES', award number 101054696. I. B. acknowledges support from the
    Deutsche Forschungsgemeinschaft under Germany’s Excellence Strategy within the Cluster of Excellence PhoenixD (EXC 2122, Project No. 390833453). S.Y. acknowledges the support of the CSF No. 202308080044. M.I. acknowledges 
    extraordinary hospitality at the Technion, especially during the week of October 8, 2023.
    
\end{acknowledgments}

\section*{Appendix\label{app:1}}

Here we describe details of our numerical simulations for the sodium atom. The calculations were done using time-dependent Schrödinger 
equation in the single active electron approximation, including ionization. The effective potential was chosen to accurately reproduce the ionization potential, as well as the key excited state energies and transition dipoles known for Na.
We used the radial potential \cite{Schweizer1999},
\begin{eqnarray}
\label{Potential}
V(r) = - \frac{1}{r} [\tilde{Z} + (Z-\tilde{Z})e^{-a_{1}r}+ a_{2}e^{-a_{3}r}]
\end{eqnarray}
with $r$ being the radial coordinate, $Z$ the atomic number, $\tilde{Z}$ the ionization stage and $a_{i}$(i=1,2,3) the potential numbers. For the sodium atom, the parameters are $Z=11$, $\tilde{Z} = 1$, $a_1 = 7.902$, $a_2 = 23.501$ and $a_3 = 2.688$ \cite{Schweizer1999}.     
To solve the TDSE, we used the method introduced in
Ref.\cite{Patchkovskii2016}. We used a radial grid extending to 100.58
a.u., with the radial step smoothly varied between $0.04a_0$ at the
origin to $0.1a_0$ beyond $r=2a_0$, $a_0$ the Bohr radius. This grid
provides a reasonable compromise between the accuracy and the
numerical efficiency of the TDSE solution (see Table \ref{tab:Na}). We
employed a uniform time step of 0.004 a.u. to ensure stable time
evolution at all field intensities of interest. We inserted a
transmission-free absorbing boundary \cite{Manolopoulos} starting at
67.88 a.u. and extending to the end of the box, which guarantees
reflection probability below $0.1\%$ for outgoing photoelectrons with
the energy above 0.54 eV \cite{Manolopoulos}. We treated the external
field in the velocity gauge and dipole approximation. Including up to
$L_{max}=20$ angular momenta was sufficient to achieve convergence for
the intensities used.
\begin{table}
  \begin{center}
  \caption{Convergence of the radial grid in TDSE calculations for Na. Ionization energies for the electronic states with the Ne core are given in eV. Only the outermost electron is listed.}
  \label{tab:Na}
    \begin{tabular}{l|ccc}
    \hline
   \multicolumn{1}{c}{State} & \multicolumn{1}{c}{Experiment$^a$} & \multicolumn{1}{c}{Reference$^b$} & \multicolumn{1}{c}{Calculated$^c$} \\
    \hline
 $3s$ (G.S.) & 5.139  & 5.139 &  5.143 \\
 $4s$ & 3.191  & 3.189 &  3.192 \\ 
 $5s$ & 4.116  & 4.115 &  4.119 \\
 \hline
 $3p$ & 2.104  & 2.105 &  2.109 \\ 
 $4p$ & 3.753  & 3.751 &  3.755 \\ 
 $5p$ & 4.345  & 4.344 &  4.348 \\ 
 \hline
 $3d$ & 3.617  & 3.616 &  3.620 \\ 
 $4d$ & 4.284  & 4.282 &  4.286 \\ 
 $5d$ & 4.592  & 4.591 &  4.595 \\ 
 \hline
 $4f$ & 4.288  & 4.289 &  4.293 \\ 
 $5f$ & 4.595  & 4.595 &  4.599 \\ 
 \hline
 $5g$ & 4.595  & 4.595 &  4.599 \\ 
    \hline
    \end{tabular}
    \begin{itemize}
        \item[$^a$] Experimental values from Ref. \cite{NIST_ASD}.
        \item[$^b$] Reference grid ($\Delta r=0.02a_0$ at the origin; $\Delta r=0.1a_0$ for  $r>4.1a_0$.
        \item[$^c$] Present grid ($\Delta r=0.04a_0$ at the origin; $\Delta r=0.1a_0$ for  $r>2a_0$, see text).
    \end{itemize}
  \end{center}
\end{table}


\begin{thebibliography}{59}%
\makeatletter
\providecommand \@ifxundefined [1]{%
 \@ifx{#1\undefined}
}%
\providecommand \@ifnum [1]{%
 \ifnum #1\expandafter \@firstoftwo
 \else \expandafter \@secondoftwo
 \fi
}%
\providecommand \@ifx [1]{%
 \ifx #1\expandafter \@firstoftwo
 \else \expandafter \@secondoftwo
 \fi
}%
\providecommand \natexlab [1]{#1}%
\providecommand \enquote  [1]{``#1''}%
\providecommand \bibnamefont  [1]{#1}%
\providecommand \bibfnamefont [1]{#1}%
\providecommand \citenamefont [1]{#1}%
\providecommand \href@noop [0]{\@secondoftwo}%
\providecommand \href [0]{\begingroup \@sanitize@url \@href}%
\providecommand \@href[1]{\@@startlink{#1}\@@href}%
\providecommand \@@href[1]{\endgroup#1\@@endlink}%
\providecommand \@sanitize@url [0]{\catcode `\\12\catcode `\$12\catcode
  `\&12\catcode `\#12\catcode `\^12\catcode `\_12\catcode `\%12\relax}%
\providecommand \@@startlink[1]{}%
\providecommand \@@endlink[0]{}%
\providecommand \url  [0]{\begingroup\@sanitize@url \@url }%
\providecommand \@url [1]{\endgroup\@href {#1}{\urlprefix }}%
\providecommand \urlprefix  [0]{URL }%
\providecommand \Eprint [0]{\href }%
\providecommand \doibase [0]{https://doi.org/}%
\providecommand \selectlanguage [0]{\@gobble}%
\providecommand \bibinfo  [0]{\@secondoftwo}%
\providecommand \bibfield  [0]{\@secondoftwo}%
\providecommand \translation [1]{[#1]}%
\providecommand \BibitemOpen [0]{}%
\providecommand \bibitemStop [0]{}%
\providecommand \bibitemNoStop [0]{.\EOS\space}%
\providecommand \EOS [0]{\spacefactor3000\relax}%
\providecommand \BibitemShut  [1]{\csname bibitem#1\endcsname}%
\let\auto@bib@innerbib\@empty
\bibitem [{\citenamefont {Ferray}\ \emph {et~al.}(1988)\citenamefont {Ferray},
  \citenamefont {Li}, \citenamefont {Lompre}, \citenamefont {Mainfray},\ and\
  \citenamefont {Manus}}]{ferray1988multiple}%
  \BibitemOpen
  \bibfield  {author} {\bibinfo {author} {\bibfnamefont {M.}~\bibnamefont
  {Ferray}}, \bibinfo {author} {\bibfnamefont {X.}~\bibnamefont {Li}}, \bibinfo
  {author} {\bibfnamefont {L.}~\bibnamefont {Lompre}}, \bibinfo {author}
  {\bibfnamefont {G.}~\bibnamefont {Mainfray}},\ and\ \bibinfo {author}
  {\bibfnamefont {C.}~\bibnamefont {Manus}},\ }\bibfield  {title} {\bibinfo
  {title} {Multiple-harmonic conversion of 1064 nm radiation in rare gases},\
  }\href {https://dx.doi.org/10.1088/0953-4075/21/3/001} {\bibfield  {journal}
  {\bibinfo  {journal} {J. Phys. B: At. Mol. Opt. Phys.}\ }\textbf {\bibinfo
  {volume} {21}},\ \bibinfo {pages} {L31} (\bibinfo {year} {1988})}\BibitemShut
  {NoStop}%
\bibitem [{\citenamefont {Drescher}\ \emph {et~al.}(2001)\citenamefont
  {Drescher}, \citenamefont {Hentschel}, \citenamefont {Kienberger},
  \citenamefont {Tempea}, \citenamefont {Spielmann}, \citenamefont {Reider},
  \citenamefont {Corkum},\ and\ \citenamefont {Krausz}}]{drescher01}%
  \BibitemOpen
  \bibfield  {author} {\bibinfo {author} {\bibfnamefont {M.}~\bibnamefont
  {Drescher}}, \bibinfo {author} {\bibfnamefont {M.}~\bibnamefont {Hentschel}},
  \bibinfo {author} {\bibfnamefont {R.}~\bibnamefont {Kienberger}}, \bibinfo
  {author} {\bibfnamefont {G.}~\bibnamefont {Tempea}}, \bibinfo {author}
  {\bibfnamefont {C.}~\bibnamefont {Spielmann}}, \bibinfo {author}
  {\bibfnamefont {G.~A.}\ \bibnamefont {Reider}}, \bibinfo {author}
  {\bibfnamefont {P.~B.}\ \bibnamefont {Corkum}},\ and\ \bibinfo {author}
  {\bibfnamefont {F.}~\bibnamefont {Krausz}},\ }\bibfield  {title} {\bibinfo
  {title} {X-ray pulses approaching the attosecond frontier},\ }\href
  {https://doi.org/10.1126/science.1058561} {\bibfield  {journal} {\bibinfo
  {journal} {Science}\ }\textbf {\bibinfo {volume} {291}},\ \bibinfo {pages}
  {1923} (\bibinfo {year} {2001})}\BibitemShut {NoStop}%
\bibitem [{\citenamefont {Paul}\ \emph {et~al.}(2001)\citenamefont {Paul},
  \citenamefont {Toma}, \citenamefont {Breger}, \citenamefont {Mullot},
  \citenamefont {Aug{\'e}}, \citenamefont {Balcou}, \citenamefont {Muller},\
  and\ \citenamefont {Agostini}}]{paul01}%
  \BibitemOpen
  \bibfield  {author} {\bibinfo {author} {\bibfnamefont {P.-M.}\ \bibnamefont
  {Paul}}, \bibinfo {author} {\bibfnamefont {E.~S.}\ \bibnamefont {Toma}},
  \bibinfo {author} {\bibfnamefont {P.}~\bibnamefont {Breger}}, \bibinfo
  {author} {\bibfnamefont {G.}~\bibnamefont {Mullot}}, \bibinfo {author}
  {\bibfnamefont {F.}~\bibnamefont {Aug{\'e}}}, \bibinfo {author}
  {\bibfnamefont {P.}~\bibnamefont {Balcou}}, \bibinfo {author} {\bibfnamefont
  {H.~G.}\ \bibnamefont {Muller}},\ and\ \bibinfo {author} {\bibfnamefont
  {P.}~\bibnamefont {Agostini}},\ }\bibfield  {title} {\bibinfo {title}
  {Observation of a train of attosecond pulses from high harmonic generation},\
  }\href {https://doi.org/10.1126/science.1058561} {\bibfield  {journal}
  {\bibinfo  {journal} {Science}\ }\textbf {\bibinfo {volume} {292}},\ \bibinfo
  {pages} {1689} (\bibinfo {year} {2001})}\BibitemShut {NoStop}%
\bibitem [{\citenamefont {Sukiasyan}\ \emph {et~al.}(2010)\citenamefont
  {Sukiasyan}, \citenamefont {Patchkovskii}, \citenamefont {Smirnova},
  \citenamefont {Brabec},\ and\ \citenamefont
  {Ivanov}}]{sukiasyan2010exchange}%
  \BibitemOpen
  \bibfield  {author} {\bibinfo {author} {\bibfnamefont {S.}~\bibnamefont
  {Sukiasyan}}, \bibinfo {author} {\bibfnamefont {S.}~\bibnamefont
  {Patchkovskii}}, \bibinfo {author} {\bibfnamefont {O.}~\bibnamefont
  {Smirnova}}, \bibinfo {author} {\bibfnamefont {T.}~\bibnamefont {Brabec}},\
  and\ \bibinfo {author} {\bibfnamefont {M.~Y.}\ \bibnamefont {Ivanov}},\
  }\bibfield  {title} {\bibinfo {title} {Exchange and polarization effect in
  high-order harmonic imaging of molecular structures},\ }\href
  {https://link.aps.org/doi/10.1103/PhysRevA.82.043414} {\bibfield  {journal}
  {\bibinfo  {journal} {Phys. Rev. A}\ }\textbf {\bibinfo {volume} {82}},\
  \bibinfo {pages} {043414} (\bibinfo {year} {2010})}\BibitemShut {NoStop}%
\bibitem [{\citenamefont {Krausz}\ and\ \citenamefont
  {Ivanov}(2009)}]{Frausz2009}%
  \BibitemOpen
  \bibfield  {author} {\bibinfo {author} {\bibfnamefont {F.}~\bibnamefont
  {Krausz}}\ and\ \bibinfo {author} {\bibfnamefont {M.}~\bibnamefont
  {Ivanov}},\ }\bibfield  {title} {\bibinfo {title} {Attosecond physics},\
  }\href {https://doi.org/10.1103/RevModPhys.81.163} {\bibfield  {journal}
  {\bibinfo  {journal} {Rev. Mod. Phys. 81, 163}\ } (\bibinfo {year}
  {2009})}\BibitemShut {NoStop}%
\bibitem [{\citenamefont {Ghimire}\ \emph {et~al.}(2011)\citenamefont
  {Ghimire}, \citenamefont {DiChiara}, \citenamefont {Sistrunk}, \citenamefont
  {Agostini}, \citenamefont {DiMauro},\ and\ \citenamefont {Reis}}]{ghimire11}%
  \BibitemOpen
  \bibfield  {author} {\bibinfo {author} {\bibfnamefont {S.}~\bibnamefont
  {Ghimire}}, \bibinfo {author} {\bibfnamefont {A.~D.}\ \bibnamefont
  {DiChiara}}, \bibinfo {author} {\bibfnamefont {E.}~\bibnamefont {Sistrunk}},
  \bibinfo {author} {\bibfnamefont {P.}~\bibnamefont {Agostini}}, \bibinfo
  {author} {\bibfnamefont {L.~F.}\ \bibnamefont {DiMauro}},\ and\ \bibinfo
  {author} {\bibfnamefont {D.~A.}\ \bibnamefont {Reis}},\ }\bibfield  {title}
  {\bibinfo {title} {Observation of high-order harmonic generation in a bulk
  crystal},\ }\href {https://doi.org/10.1038/nphys1847} {\bibfield  {journal}
  {\bibinfo  {journal} {Nat. Phys.}\ }\textbf {\bibinfo {volume} {7}},\
  \bibinfo {pages} {138} (\bibinfo {year} {2011})}\BibitemShut {NoStop}%
\bibitem [{\citenamefont {Krausz}\ and\ \citenamefont
  {Stockman}(2014)}]{krausz14}%
  \BibitemOpen
  \bibfield  {author} {\bibinfo {author} {\bibfnamefont {F.}~\bibnamefont
  {Krausz}}\ and\ \bibinfo {author} {\bibfnamefont {M.~I.}\ \bibnamefont
  {Stockman}},\ }\bibfield  {title} {\bibinfo {title} {Attosecond metrology:
  from electron capture to future signal processing},\ }\href
  {https://doi.org/10.1038/35107000} {\bibfield  {journal} {\bibinfo  {journal}
  {Nat. Photon.}\ }\textbf {\bibinfo {volume} {8}},\ \bibinfo {pages} {205}
  (\bibinfo {year} {2014})}\BibitemShut {NoStop}%
\bibitem [{\citenamefont {Vampa}\ \emph {et~al.}(2014)\citenamefont {Vampa},
  \citenamefont {McDonald}, \citenamefont {Orlando}, \citenamefont {Klug},
  \citenamefont {Corkum},\ and\ \citenamefont {Brabec}}]{vampa14}%
  \BibitemOpen
  \bibfield  {author} {\bibinfo {author} {\bibfnamefont {G.}~\bibnamefont
  {Vampa}}, \bibinfo {author} {\bibfnamefont {C.~R.}\ \bibnamefont {McDonald}},
  \bibinfo {author} {\bibfnamefont {G.}~\bibnamefont {Orlando}}, \bibinfo
  {author} {\bibfnamefont {D.~D.}\ \bibnamefont {Klug}}, \bibinfo {author}
  {\bibfnamefont {P.~B.}\ \bibnamefont {Corkum}},\ and\ \bibinfo {author}
  {\bibfnamefont {T.}~\bibnamefont {Brabec}},\ }\bibfield  {title} {\bibinfo
  {title} {Theoretical analysis of high-harmonic generation in solids},\ }\href
  {https://doi.org/10.1103/PhysRevLett.113.073901} {\bibfield  {journal}
  {\bibinfo  {journal} {Phys. Rev. Lett.}\ }\textbf {\bibinfo {volume} {113}},\
  \bibinfo {pages} {073901} (\bibinfo {year} {2014})}\BibitemShut {NoStop}%
\bibitem [{\citenamefont {Liu}\ \emph {et~al.}(2017)\citenamefont {Liu},
  \citenamefont {Li}, \citenamefont {You}, \citenamefont {Ghimire},
  \citenamefont {Heinz},\ and\ \citenamefont {Reis}}]{liu17}%
  \BibitemOpen
  \bibfield  {author} {\bibinfo {author} {\bibfnamefont {H.}~\bibnamefont
  {Liu}}, \bibinfo {author} {\bibfnamefont {Y.}~\bibnamefont {Li}}, \bibinfo
  {author} {\bibfnamefont {Y.~S.}\ \bibnamefont {You}}, \bibinfo {author}
  {\bibfnamefont {S.}~\bibnamefont {Ghimire}}, \bibinfo {author} {\bibfnamefont
  {T.~F.}\ \bibnamefont {Heinz}},\ and\ \bibinfo {author} {\bibfnamefont
  {D.~A.}\ \bibnamefont {Reis}},\ }\bibfield  {title} {\bibinfo {title}
  {High-harmonic generation from an atomically thin semiconductor},\
  }\href@noop {} {\bibfield  {journal} {\bibinfo  {journal} {Nature Physics}\
  }\textbf {\bibinfo {volume} {13}},\ \bibinfo {pages} {262} (\bibinfo {year}
  {2017})}\BibitemShut {NoStop}%
\bibitem [{\citenamefont {Ghimire}\ and\ \citenamefont
  {Reis}(2019)}]{ghimire19}%
  \BibitemOpen
  \bibfield  {author} {\bibinfo {author} {\bibfnamefont {S.}~\bibnamefont
  {Ghimire}}\ and\ \bibinfo {author} {\bibfnamefont {D.~A.}\ \bibnamefont
  {Reis}},\ }\bibfield  {title} {\bibinfo {title} {High-harmonic generation
  from solids},\ }\href@noop {} {\bibfield  {journal} {\bibinfo  {journal}
  {Nature physics}\ }\textbf {\bibinfo {volume} {15}},\ \bibinfo {pages} {10}
  (\bibinfo {year} {2019})}\BibitemShut {NoStop}%
\bibitem [{\citenamefont {Goulielmakis}\ and\ \citenamefont
  {Brabec}(2022)}]{goulielmakis22}%
  \BibitemOpen
  \bibfield  {author} {\bibinfo {author} {\bibfnamefont {E.}~\bibnamefont
  {Goulielmakis}}\ and\ \bibinfo {author} {\bibfnamefont {T.}~\bibnamefont
  {Brabec}},\ }\bibfield  {title} {\bibinfo {title} {High harmonic generation
  in condensed matter},\ }\href@noop {} {\bibfield  {journal} {\bibinfo
  {journal} {Nature Photonics}\ }\textbf {\bibinfo {volume} {16}},\ \bibinfo
  {pages} {411} (\bibinfo {year} {2022})}\BibitemShut {NoStop}%
\bibitem [{\citenamefont {Lewenstein}\ \emph {et~al.}(2021)\citenamefont
  {Lewenstein}, \citenamefont {Ciappina}, \citenamefont {Pisanty},
  \citenamefont {Rivera-Dean}, \citenamefont {Stammer}, \citenamefont
  {Lamprou},\ and\ \citenamefont {Tzallas}}]{lewenstein2021generation}%
  \BibitemOpen
  \bibfield  {author} {\bibinfo {author} {\bibfnamefont {M.}~\bibnamefont
  {Lewenstein}}, \bibinfo {author} {\bibfnamefont {M.}~\bibnamefont
  {Ciappina}}, \bibinfo {author} {\bibfnamefont {E.}~\bibnamefont {Pisanty}},
  \bibinfo {author} {\bibfnamefont {J.}~\bibnamefont {Rivera-Dean}}, \bibinfo
  {author} {\bibfnamefont {P.}~\bibnamefont {Stammer}}, \bibinfo {author}
  {\bibfnamefont {T.}~\bibnamefont {Lamprou}},\ and\ \bibinfo {author}
  {\bibfnamefont {P.}~\bibnamefont {Tzallas}},\ }\bibfield  {title} {\bibinfo
  {title} {Generation of optical {Schr{\"o}dinger} cat states in intense
  laser-matter interactions},\ }\href
  {https://doi.org/10.1038/s41567-021-01317-w} {\bibfield  {journal} {\bibinfo
  {journal} {Nat. Phys.}\ }\textbf {\bibinfo {volume} {17}},\ \bibinfo {pages}
  {1104} (\bibinfo {year} {2021})}\BibitemShut {NoStop}%
\bibitem [{\citenamefont {Tsatrafyllis}\ \emph {et~al.}(2017)\citenamefont
  {Tsatrafyllis}, \citenamefont {Kominis}, \citenamefont {Gonoskov},\ and\
  \citenamefont {Tzallas}}]{tsatrafyllis17}%
  \BibitemOpen
  \bibfield  {author} {\bibinfo {author} {\bibfnamefont {N.}~\bibnamefont
  {Tsatrafyllis}}, \bibinfo {author} {\bibfnamefont {I.}~\bibnamefont
  {Kominis}}, \bibinfo {author} {\bibfnamefont {I.}~\bibnamefont {Gonoskov}},\
  and\ \bibinfo {author} {\bibfnamefont {P.}~\bibnamefont {Tzallas}},\
  }\bibfield  {title} {\bibinfo {title} {High-order harmonics measured by the
  photon statistics of the infrared driving-field exiting the atomic medium},\
  }\href {https://doi.org/10.1038/ncomms15170} {\bibfield  {journal} {\bibinfo
  {journal} {Nat. Commun.}\ }\textbf {\bibinfo {volume} {8}},\ \bibinfo {pages}
  {15170} (\bibinfo {year} {2017})}\BibitemShut {NoStop}%
\bibitem [{\citenamefont {Gorlach}\ \emph {et~al.}(2020)\citenamefont
  {Gorlach}, \citenamefont {Neufeld}, \citenamefont {Rivera}, \citenamefont
  {Cohen},\ and\ \citenamefont {Kaminer}}]{gorlach20}%
  \BibitemOpen
  \bibfield  {author} {\bibinfo {author} {\bibfnamefont {A.}~\bibnamefont
  {Gorlach}}, \bibinfo {author} {\bibfnamefont {O.}~\bibnamefont {Neufeld}},
  \bibinfo {author} {\bibfnamefont {N.}~\bibnamefont {Rivera}}, \bibinfo
  {author} {\bibfnamefont {O.}~\bibnamefont {Cohen}},\ and\ \bibinfo {author}
  {\bibfnamefont {I.}~\bibnamefont {Kaminer}},\ }\bibfield  {title} {\bibinfo
  {title} {The quantum-optical nature of high harmonic generation},\ }\href
  {https://doi.org/10.1038/s41467-020-18218-w} {\bibfield  {journal} {\bibinfo
  {journal} {Nat. Commun.}\ }\textbf {\bibinfo {volume} {11}},\ \bibinfo
  {pages} {4598} (\bibinfo {year} {2020})}\BibitemShut {NoStop}%
\bibitem [{\citenamefont {Rivera-Dean}\ \emph
  {et~al.}(2022{\natexlab{a}})\citenamefont {Rivera-Dean}, \citenamefont
  {Lamprou}, \citenamefont {Pisanty}, \citenamefont {Stammer}, \citenamefont
  {Ord\'o\~nez}, \citenamefont {Maxwell}, \citenamefont {Ciappina},
  \citenamefont {Lewenstein},\ and\ \citenamefont {Tzallas}}]{rivera-dean22}%
  \BibitemOpen
  \bibfield  {author} {\bibinfo {author} {\bibfnamefont {J.}~\bibnamefont
  {Rivera-Dean}}, \bibinfo {author} {\bibfnamefont {T.}~\bibnamefont
  {Lamprou}}, \bibinfo {author} {\bibfnamefont {E.}~\bibnamefont {Pisanty}},
  \bibinfo {author} {\bibfnamefont {P.}~\bibnamefont {Stammer}}, \bibinfo
  {author} {\bibfnamefont {A.~F.}\ \bibnamefont {Ord\'o\~nez}}, \bibinfo
  {author} {\bibfnamefont {A.~S.}\ \bibnamefont {Maxwell}}, \bibinfo {author}
  {\bibfnamefont {M.~F.}\ \bibnamefont {Ciappina}}, \bibinfo {author}
  {\bibfnamefont {M.}~\bibnamefont {Lewenstein}},\ and\ \bibinfo {author}
  {\bibfnamefont {P.}~\bibnamefont {Tzallas}},\ }\bibfield  {title} {\bibinfo
  {title} {Strong laser fields and their power to generate controllable
  high-photon-number coherent-state superpositions},\ }\href
  {https://doi/10.1103/PhysRevA.105.033714} {\bibfield  {journal} {\bibinfo
  {journal} {Phys. Rev. A}\ }\textbf {\bibinfo {volume} {105}},\ \bibinfo
  {pages} {033714} (\bibinfo {year} {2022}{\natexlab{a}})}\BibitemShut
  {NoStop}%
\bibitem [{\citenamefont {Rivera-Dean}\ \emph
  {et~al.}(2022{\natexlab{b}})\citenamefont {Rivera-Dean}, \citenamefont
  {Stammer}, \citenamefont {Maxwell}, \citenamefont {Lamprou}, \citenamefont
  {Tzallas}, \citenamefont {Lewenstein},\ and\ \citenamefont
  {Ciappina}}]{rivera-dean22a}%
  \BibitemOpen
  \bibfield  {author} {\bibinfo {author} {\bibfnamefont {J.}~\bibnamefont
  {Rivera-Dean}}, \bibinfo {author} {\bibfnamefont {P.}~\bibnamefont
  {Stammer}}, \bibinfo {author} {\bibfnamefont {A.~S.}\ \bibnamefont
  {Maxwell}}, \bibinfo {author} {\bibfnamefont {T.}~\bibnamefont {Lamprou}},
  \bibinfo {author} {\bibfnamefont {P.}~\bibnamefont {Tzallas}}, \bibinfo
  {author} {\bibfnamefont {M.}~\bibnamefont {Lewenstein}},\ and\ \bibinfo
  {author} {\bibfnamefont {M.~F.}\ \bibnamefont {Ciappina}},\ }\bibfield
  {title} {\bibinfo {title} {Light-matter entanglement after above-threshold
  ionization processes in atoms},\ }\href
  {https://doi.org/10.1103/PhysRevA.106.063705} {\bibfield  {journal} {\bibinfo
   {journal} {Phys. Rev. A}\ }\textbf {\bibinfo {volume} {106}},\ \bibinfo
  {pages} {063705} (\bibinfo {year} {2022}{\natexlab{b}})}\BibitemShut
  {NoStop}%
\bibitem [{\citenamefont {Maxwell}\ \emph {et~al.}(2022)\citenamefont
  {Maxwell}, \citenamefont {Madsen},\ and\ \citenamefont
  {Lewenstein}}]{maxwell22}%
  \BibitemOpen
  \bibfield  {author} {\bibinfo {author} {\bibfnamefont {A.~S.}\ \bibnamefont
  {Maxwell}}, \bibinfo {author} {\bibfnamefont {L.~B.}\ \bibnamefont
  {Madsen}},\ and\ \bibinfo {author} {\bibfnamefont {M.}~\bibnamefont
  {Lewenstein}},\ }\bibfield  {title} {\bibinfo {title} {Entanglement of
  orbital angular momentum in non-sequential double ionization},\ }\href
  {https://doi.org/10.1038/s41467-022-32128-z} {\bibfield  {journal} {\bibinfo
  {journal} {Nat. Commun.}\ }\textbf {\bibinfo {volume} {13}},\ \bibinfo
  {pages} {4706} (\bibinfo {year} {2022})}\BibitemShut {NoStop}%
\bibitem [{\citenamefont {Stammer}(2022)}]{stammer22}%
  \BibitemOpen
  \bibfield  {author} {\bibinfo {author} {\bibfnamefont {P.}~\bibnamefont
  {Stammer}},\ }\bibfield  {title} {\bibinfo {title} {Theory of entanglement
  and measurement in high-order harmonic generation},\ }\href
  {https://doi.org/10.1103/PhysRevA.106.L050402} {\bibfield  {journal}
  {\bibinfo  {journal} {Phys. Rev. A}\ }\textbf {\bibinfo {volume} {106}},\
  \bibinfo {pages} {L050402} (\bibinfo {year} {2022})}\BibitemShut {NoStop}%
\bibitem [{\citenamefont {Stammer}\ \emph {et~al.}(2022)\citenamefont
  {Stammer}, \citenamefont {Rivera-Dean}, \citenamefont {Lamprou},
  \citenamefont {Pisanty}, \citenamefont {Ciappina}, \citenamefont {Tzallas},\
  and\ \citenamefont {Lewenstein}}]{stammer22a}%
  \BibitemOpen
  \bibfield  {author} {\bibinfo {author} {\bibfnamefont {P.}~\bibnamefont
  {Stammer}}, \bibinfo {author} {\bibfnamefont {J.}~\bibnamefont
  {Rivera-Dean}}, \bibinfo {author} {\bibfnamefont {T.}~\bibnamefont
  {Lamprou}}, \bibinfo {author} {\bibfnamefont {E.}~\bibnamefont {Pisanty}},
  \bibinfo {author} {\bibfnamefont {M.~F.}\ \bibnamefont {Ciappina}}, \bibinfo
  {author} {\bibfnamefont {P.}~\bibnamefont {Tzallas}},\ and\ \bibinfo {author}
  {\bibfnamefont {M.}~\bibnamefont {Lewenstein}},\ }\bibfield  {title}
  {\bibinfo {title} {High photon number entangled states and coherent state
  superposition from the extreme ultraviolet to the far infrared},\ }\href
  {doi.org/10.1103/PhysRevA.106.063705} {\bibfield  {journal} {\bibinfo
  {journal} {Phys. Rev. Lett.}\ }\textbf {\bibinfo {volume} {128}},\ \bibinfo
  {pages} {123603} (\bibinfo {year} {2022})}\BibitemShut {NoStop}%
\bibitem [{\citenamefont {Stammer}\ \emph {et~al.}(2023)\citenamefont
  {Stammer}, \citenamefont {Rivera-Dean}, \citenamefont {Maxwell},
  \citenamefont {Lamprou}, \citenamefont {Ord\'o\~nez}, \citenamefont
  {Ciappina}, \citenamefont {Tzallas},\ and\ \citenamefont
  {Lewenstein}}]{Lewenstein2023}%
  \BibitemOpen
  \bibfield  {author} {\bibinfo {author} {\bibfnamefont {P.}~\bibnamefont
  {Stammer}}, \bibinfo {author} {\bibfnamefont {J.}~\bibnamefont
  {Rivera-Dean}}, \bibinfo {author} {\bibfnamefont {A.}~\bibnamefont
  {Maxwell}}, \bibinfo {author} {\bibfnamefont {T.}~\bibnamefont {Lamprou}},
  \bibinfo {author} {\bibfnamefont {A.}~\bibnamefont {Ord\'o\~nez}}, \bibinfo
  {author} {\bibfnamefont {M.~F.}\ \bibnamefont {Ciappina}}, \bibinfo {author}
  {\bibfnamefont {P.}~\bibnamefont {Tzallas}},\ and\ \bibinfo {author}
  {\bibfnamefont {M.}~\bibnamefont {Lewenstein}},\ }\bibfield  {title}
  {\bibinfo {title} {Quantum electrodynamics of intense laser-matter
  interactions: A tool for quantum state engineering},\ }\href
  {https://doi.org/10.1103/PRXQuantum.4.010201} {\bibfield  {journal} {\bibinfo
   {journal} {PRX Quantum 4, 010201}\ } (\bibinfo {year} {2023})}\BibitemShut
  {NoStop}%
\bibitem [{\citenamefont {Bhattacharya}\ \emph {et~al.}(2023)\citenamefont
  {Bhattacharya}, \citenamefont {Lamprou}, \citenamefont {Maxwell},
  \citenamefont {Ordonez}, \citenamefont {Pisanty}, \citenamefont
  {Rivera-Dean}, \citenamefont {Stammer}, \citenamefont {Ciappina},
  \citenamefont {Lewenstein},\ and\ \citenamefont {Tzallas}}]{bhattacharya23}%
  \BibitemOpen
  \bibfield  {author} {\bibinfo {author} {\bibfnamefont {U.}~\bibnamefont
  {Bhattacharya}}, \bibinfo {author} {\bibfnamefont {T.}~\bibnamefont
  {Lamprou}}, \bibinfo {author} {\bibfnamefont {A.~S.}\ \bibnamefont
  {Maxwell}}, \bibinfo {author} {\bibfnamefont {A.}~\bibnamefont {Ordonez}},
  \bibinfo {author} {\bibfnamefont {E.}~\bibnamefont {Pisanty}}, \bibinfo
  {author} {\bibfnamefont {J.}~\bibnamefont {Rivera-Dean}}, \bibinfo {author}
  {\bibfnamefont {P.}~\bibnamefont {Stammer}}, \bibinfo {author} {\bibfnamefont
  {M.~F.}\ \bibnamefont {Ciappina}}, \bibinfo {author} {\bibfnamefont
  {M.}~\bibnamefont {Lewenstein}},\ and\ \bibinfo {author} {\bibfnamefont
  {P.}~\bibnamefont {Tzallas}},\ }\bibfield  {title} {\bibinfo {title}
  {Strong--laser--field physics, non--classical light states and quantum
  information science},\ }\href {https://doi.org/10.1088/1361-6633} {\bibfield
  {journal} {\bibinfo  {journal} {Rep. Prog. Phys.}\ }\textbf {\bibinfo
  {volume} {86}} (\bibinfo {year} {2023})}\BibitemShut {NoStop}%
\bibitem [{\citenamefont {Lemieux}\ \emph {et~al.}(2024)\citenamefont
  {Lemieux}, \citenamefont {Jalil}, \citenamefont {Purschke}, \citenamefont
  {Boroumand}, \citenamefont {Villeneuve}, \citenamefont {Naumov},
  \citenamefont {Brabec},\ and\ \citenamefont {Vampa}}]{lemieux2024photon}%
  \BibitemOpen
  \bibfield  {author} {\bibinfo {author} {\bibfnamefont {S.}~\bibnamefont
  {Lemieux}}, \bibinfo {author} {\bibfnamefont {S.~A.}\ \bibnamefont {Jalil}},
  \bibinfo {author} {\bibfnamefont {D.}~\bibnamefont {Purschke}}, \bibinfo
  {author} {\bibfnamefont {N.}~\bibnamefont {Boroumand}}, \bibinfo {author}
  {\bibfnamefont {D.}~\bibnamefont {Villeneuve}}, \bibinfo {author}
  {\bibfnamefont {A.}~\bibnamefont {Naumov}}, \bibinfo {author} {\bibfnamefont
  {T.}~\bibnamefont {Brabec}},\ and\ \bibinfo {author} {\bibfnamefont
  {G.}~\bibnamefont {Vampa}},\ }\bibfield  {title} {\bibinfo {title} {Photon
  bunching in high-harmonic emission controlled by quantum light},\ }\href
  {https://doi.org/10.48550/arXiv.2404.05474} {\bibfield  {journal} {\bibinfo
  {journal} {arXiv preprint arXiv:2404.05474}\ } (\bibinfo {year}
  {2024})}\BibitemShut {NoStop}%
\bibitem [{\citenamefont {Theidel}\ \emph {et~al.}(2024)\citenamefont
  {Theidel}, \citenamefont {Cotte}, \citenamefont {Sondenheimer}, \citenamefont
  {Shiriaeva}, \citenamefont {Froidevaux}, \citenamefont {Severin},
  \citenamefont {Mosel}, \citenamefont {Merdji-Larue}, \citenamefont
  {Fr{\"o}hlich}, \citenamefont {Weber} \emph {et~al.}}]{theidel2024evidence}%
  \BibitemOpen
  \bibfield  {author} {\bibinfo {author} {\bibfnamefont {D.}~\bibnamefont
  {Theidel}}, \bibinfo {author} {\bibfnamefont {V.}~\bibnamefont {Cotte}},
  \bibinfo {author} {\bibfnamefont {R.}~\bibnamefont {Sondenheimer}}, \bibinfo
  {author} {\bibfnamefont {V.}~\bibnamefont {Shiriaeva}}, \bibinfo {author}
  {\bibfnamefont {M.}~\bibnamefont {Froidevaux}}, \bibinfo {author}
  {\bibfnamefont {V.}~\bibnamefont {Severin}}, \bibinfo {author} {\bibfnamefont
  {P.}~\bibnamefont {Mosel}}, \bibinfo {author} {\bibfnamefont
  {A.}~\bibnamefont {Merdji-Larue}}, \bibinfo {author} {\bibfnamefont
  {S.}~\bibnamefont {Fr{\"o}hlich}}, \bibinfo {author} {\bibfnamefont {K.-A.}\
  \bibnamefont {Weber}}, \emph {et~al.},\ }\bibfield  {title} {\bibinfo {title}
  {Evidence of the quantum-optical nature of high-harmonic generation},\ }\href
  {https://doi.org/10.48550/arXiv.2405.15022} {\bibfield  {journal} {\bibinfo
  {journal} {arXiv preprint arXiv:2405.15022}\ } (\bibinfo {year}
  {2024})}\BibitemShut {NoStop}%
\bibitem [{\citenamefont {Rasputnyi}\ \emph {et~al.}(2024)\citenamefont
  {Rasputnyi}, \citenamefont {Chen}, \citenamefont {Birk}, \citenamefont
  {Cohen}, \citenamefont {Kaminer}, \citenamefont {Kr{\"u}ger}, \citenamefont
  {Seletskiy}, \citenamefont {Chekhova},\ and\ \citenamefont
  {Tani}}]{rasputnyi2024high}%
  \BibitemOpen
  \bibfield  {author} {\bibinfo {author} {\bibfnamefont {A.}~\bibnamefont
  {Rasputnyi}}, \bibinfo {author} {\bibfnamefont {Z.}~\bibnamefont {Chen}},
  \bibinfo {author} {\bibfnamefont {M.}~\bibnamefont {Birk}}, \bibinfo {author}
  {\bibfnamefont {O.}~\bibnamefont {Cohen}}, \bibinfo {author} {\bibfnamefont
  {I.}~\bibnamefont {Kaminer}}, \bibinfo {author} {\bibfnamefont
  {M.}~\bibnamefont {Kr{\"u}ger}}, \bibinfo {author} {\bibfnamefont
  {D.}~\bibnamefont {Seletskiy}}, \bibinfo {author} {\bibfnamefont
  {M.}~\bibnamefont {Chekhova}},\ and\ \bibinfo {author} {\bibfnamefont
  {F.}~\bibnamefont {Tani}},\ }\bibfield  {title} {\bibinfo {title} {High
  harmonic generation by bright squeezed vacuum},\ }\href
  {https://doi.org/10.48550/arXiv.2403.15337} {\bibfield  {journal} {\bibinfo
  {journal} {arXiv preprint arXiv:2403.15337}\ } (\bibinfo {year}
  {2024})}\BibitemShut {NoStop}%
\bibitem [{\citenamefont {Sundaram}\ and\ \citenamefont
  {Milonni}(1990)}]{sundaram90}%
  \BibitemOpen
  \bibfield  {author} {\bibinfo {author} {\bibfnamefont {B.}~\bibnamefont
  {Sundaram}}\ and\ \bibinfo {author} {\bibfnamefont {P.~W.}\ \bibnamefont
  {Milonni}},\ }\bibfield  {title} {\bibinfo {title} {High-order harmonic
  generation: Simplified model and relevance of single-atom theories to
  experiment},\ }\href {https://doi.org/10.1103/PhysRevA.41.6571} {\bibfield
  {journal} {\bibinfo  {journal} {Phys. Rev. A}\ }\textbf {\bibinfo {volume}
  {41}},\ \bibinfo {pages} {6571} (\bibinfo {year} {1990})}\BibitemShut
  {NoStop}%
\bibitem [{\citenamefont {Gauthey}\ \emph {et~al.}(1995)\citenamefont
  {Gauthey}, \citenamefont {Keitel}, \citenamefont {Knight},\ and\
  \citenamefont {Maquet}}]{gauthey95}%
  \BibitemOpen
  \bibfield  {author} {\bibinfo {author} {\bibfnamefont {F.~I.}\ \bibnamefont
  {Gauthey}}, \bibinfo {author} {\bibfnamefont {C.~H.}\ \bibnamefont {Keitel}},
  \bibinfo {author} {\bibfnamefont {P.~L.}\ \bibnamefont {Knight}},\ and\
  \bibinfo {author} {\bibfnamefont {A.}~\bibnamefont {Maquet}},\ }\bibfield
  {title} {\bibinfo {title} {Role of initial coherence in the generation of
  harmonics and sidebands from a strongly driven two-level atom},\ }\href
  {https://doi.org/10.1103/PhysRevA.52.525} {\bibfield  {journal} {\bibinfo
  {journal} {Phys. Rev. A}\ }\textbf {\bibinfo {volume} {52}},\ \bibinfo
  {pages} {525} (\bibinfo {year} {1995})}\BibitemShut {NoStop}%
\bibitem [{\citenamefont {Gonoskov}\ \emph {et~al.}(2016)\citenamefont
  {Gonoskov}, \citenamefont {Tsatrafyllis},\ and\ \citenamefont
  {Kominis}}]{gonoskov2016quantum}%
  \BibitemOpen
  \bibfield  {author} {\bibinfo {author} {\bibfnamefont {I.}~\bibnamefont
  {Gonoskov}}, \bibinfo {author} {\bibfnamefont {N.}~\bibnamefont
  {Tsatrafyllis}},\ and\ \bibinfo {author} {\bibfnamefont {I.}~\bibnamefont
  {Kominis}},\ }\bibfield  {title} {\bibinfo {title} {Quantum optical
  signatures in strong-field laser physics: Infrared photon counting in
  high-order-harmonic generation},\ }\href
  {https://doi.org/https://doi.org/10.1038/srep32821} {\bibfield  {journal}
  {\bibinfo  {journal} {Sci. Rep.}\ }\textbf {\bibinfo {volume} {6}},\ \bibinfo
  {pages} {32821} (\bibinfo {year} {2016})}\BibitemShut {NoStop}%
\bibitem [{\citenamefont {Pizzi}\ \emph {et~al.}(2023)\citenamefont {Pizzi},
  \citenamefont {Gorlach}, \citenamefont {Rivera}, \citenamefont {Nunnenkamp},\
  and\ \citenamefont {Kaminer}}]{pizzi23}%
  \BibitemOpen
  \bibfield  {author} {\bibinfo {author} {\bibfnamefont {A.}~\bibnamefont
  {Pizzi}}, \bibinfo {author} {\bibfnamefont {A.}~\bibnamefont {Gorlach}},
  \bibinfo {author} {\bibfnamefont {N.}~\bibnamefont {Rivera}}, \bibinfo
  {author} {\bibfnamefont {A.}~\bibnamefont {Nunnenkamp}},\ and\ \bibinfo
  {author} {\bibfnamefont {I.}~\bibnamefont {Kaminer}},\ }\bibfield  {title}
  {\bibinfo {title} {Light emission from strongly driven many-body systems},\
  }\href {https://doi.org/10.1038/s41567-022-01910-7} {\bibfield  {journal}
  {\bibinfo  {journal} {Nat. Phys.}\ }\textbf {\bibinfo {volume} {19}},\
  \bibinfo {pages} {551} (\bibinfo {year} {2023})}\BibitemShut {NoStop}%
\bibitem [{\citenamefont {Tzallas}(2023)}]{tzallas23}%
  \BibitemOpen
  \bibfield  {author} {\bibinfo {author} {\bibfnamefont {P.}~\bibnamefont
  {Tzallas}},\ }\bibfield  {title} {\bibinfo {title} {Quantum correlated atoms
  in intense laser fields},\ }\href
  {https://doi.org/10.1038/s41567-022-01928-x} {\bibfield  {journal} {\bibinfo
  {journal} {Nat. Phys.}\ }\textbf {\bibinfo {volume} {19}},\ \bibinfo {pages}
  {472} (\bibinfo {year} {2023})}\BibitemShut {NoStop}%
\bibitem [{\citenamefont {Lange}\ \emph {et~al.}(2023)\citenamefont {Lange},
  \citenamefont {Hansen},\ and\ \citenamefont {Madsen}}]{lange2023electron}%
  \BibitemOpen
  \bibfield  {author} {\bibinfo {author} {\bibfnamefont {C.~S.}\ \bibnamefont
  {Lange}}, \bibinfo {author} {\bibfnamefont {T.}~\bibnamefont {Hansen}},\ and\
  \bibinfo {author} {\bibfnamefont {L.~B.}\ \bibnamefont {Madsen}},\ }\bibfield
   {title} {\bibinfo {title} {Electron-correlation induced nonclassicallity of
  light from high-harmonic generation},\ }\href
  {https://doi.org/10.48550/arXiv.2312.08942} {\bibfield  {journal} {\bibinfo
  {journal} {arXiv preprint arXiv:2312.08942}\ } (\bibinfo {year}
  {2023})}\BibitemShut {NoStop}%
\bibitem [{\citenamefont {Lamprou}\ \emph {et~al.}(2023)\citenamefont
  {Lamprou}, \citenamefont {Rivera-Dean}, \citenamefont {Stammer},
  \citenamefont {Lewenstein},\ and\ \citenamefont
  {Tzallas}}]{lamprou2023nonlinear}%
  \BibitemOpen
  \bibfield  {author} {\bibinfo {author} {\bibfnamefont {T.}~\bibnamefont
  {Lamprou}}, \bibinfo {author} {\bibfnamefont {J.}~\bibnamefont
  {Rivera-Dean}}, \bibinfo {author} {\bibfnamefont {P.}~\bibnamefont
  {Stammer}}, \bibinfo {author} {\bibfnamefont {M.}~\bibnamefont
  {Lewenstein}},\ and\ \bibinfo {author} {\bibfnamefont {P.}~\bibnamefont
  {Tzallas}},\ }\bibfield  {title} {\bibinfo {title} {Nonlinear optics using
  intense optical schrödinger "cat" states},\ }\href
  {https://doi.org/10.48550/arXiv.2306.14480} {\bibfield  {journal} {\bibinfo
  {journal} {arXiv preprint arXiv:2306.14480}\ } (\bibinfo {year}
  {2023})}\BibitemShut {NoStop}%
\bibitem [{\citenamefont {Spasibko}\ \emph {et~al.}(2017)\citenamefont
  {Spasibko}, \citenamefont {Kopylov}, \citenamefont {Krutyanskiy},
  \citenamefont {Murzina}, \citenamefont {Leuchs},\ and\ \citenamefont
  {Chekhova}}]{Spasibko2017}%
  \BibitemOpen
  \bibfield  {author} {\bibinfo {author} {\bibfnamefont {K.~Y.}\ \bibnamefont
  {Spasibko}}, \bibinfo {author} {\bibfnamefont {D.~A.}\ \bibnamefont
  {Kopylov}}, \bibinfo {author} {\bibfnamefont {V.~L.}\ \bibnamefont
  {Krutyanskiy}}, \bibinfo {author} {\bibfnamefont {T.~V.}\ \bibnamefont
  {Murzina}}, \bibinfo {author} {\bibfnamefont {G.}~\bibnamefont {Leuchs}},\
  and\ \bibinfo {author} {\bibfnamefont {M.~V.}\ \bibnamefont {Chekhova}},\
  }\bibfield  {title} {\bibinfo {title} {Multiphoton effects enhanced due to
  ultrafast photon-number fluctuations},\ }\href
  {https://doi.org/10.1103/PhysRevLett.119.223603} {\bibfield  {journal}
  {\bibinfo  {journal} {Phys. Rev. Lett.}\ }\textbf {\bibinfo {volume} {119}},\
  \bibinfo {pages} {223603} (\bibinfo {year} {2017})}\BibitemShut {NoStop}%
\bibitem [{\citenamefont {Even~Tzur}\ \emph {et~al.}(2023)\citenamefont
  {Even~Tzur}, \citenamefont {Birk}, \citenamefont {Gorlach}, \citenamefont
  {Kr{\"u}ger}, \citenamefont {Kaminer},\ and\ \citenamefont
  {Cohen}}]{even2023photon}%
  \BibitemOpen
  \bibfield  {author} {\bibinfo {author} {\bibfnamefont {M.}~\bibnamefont
  {Even~Tzur}}, \bibinfo {author} {\bibfnamefont {M.}~\bibnamefont {Birk}},
  \bibinfo {author} {\bibfnamefont {A.}~\bibnamefont {Gorlach}}, \bibinfo
  {author} {\bibfnamefont {M.}~\bibnamefont {Kr{\"u}ger}}, \bibinfo {author}
  {\bibfnamefont {I.}~\bibnamefont {Kaminer}},\ and\ \bibinfo {author}
  {\bibfnamefont {O.}~\bibnamefont {Cohen}},\ }\bibfield  {title} {\bibinfo
  {title} {Photon-statistics force in ultrafast electron dynamics},\ }\href
  {https://doi.org/10.1038/s41566-023-01209-w} {\bibfield  {journal} {\bibinfo
  {journal} {Nat. Photon.}\ }\textbf {\bibinfo {volume} {17}},\ \bibinfo
  {pages} {501} (\bibinfo {year} {2023})}\BibitemShut {NoStop}%
\bibitem [{\citenamefont {Gorlach}\ \emph {et~al.}(2023)\citenamefont
  {Gorlach}, \citenamefont {Tzur}, \citenamefont {Birk}, \citenamefont
  {Krüger}, \citenamefont {Rivera}, \citenamefont {Cohen},\ and\ \citenamefont
  {Kaminer}}]{Kaminer2023}%
  \BibitemOpen
  \bibfield  {author} {\bibinfo {author} {\bibfnamefont {A.}~\bibnamefont
  {Gorlach}}, \bibinfo {author} {\bibfnamefont {M.~E.}\ \bibnamefont {Tzur}},
  \bibinfo {author} {\bibfnamefont {M.}~\bibnamefont {Birk}}, \bibinfo {author}
  {\bibfnamefont {M.}~\bibnamefont {Krüger}}, \bibinfo {author} {\bibfnamefont
  {N.}~\bibnamefont {Rivera}}, \bibinfo {author} {\bibfnamefont
  {O.}~\bibnamefont {Cohen}},\ and\ \bibinfo {author} {\bibfnamefont
  {I.}~\bibnamefont {Kaminer}},\ }\bibfield  {title} {\bibinfo {title}
  {High-harmonic generation driven by quantum light},\ }\href
  {https://doi.org/10.1038/s41567-023-02127-y} {\bibfield  {journal} {\bibinfo
  {journal} {Nat. Phys.}\ }\textbf {\bibinfo {volume} {19}},\ \bibinfo {pages}
  {1689–1696} (\bibinfo {year} {2023})}\BibitemShut {NoStop}%
\bibitem [{\citenamefont {Iskhakov}\ \emph {et~al.}(2009)\citenamefont
  {Iskhakov}, \citenamefont {Chekhova},\ and\ \citenamefont
  {Leuchs}}]{iskhakov2009generation}%
  \BibitemOpen
  \bibfield  {author} {\bibinfo {author} {\bibfnamefont {T.}~\bibnamefont
  {Iskhakov}}, \bibinfo {author} {\bibfnamefont {M.~V.}\ \bibnamefont
  {Chekhova}},\ and\ \bibinfo {author} {\bibfnamefont {G.}~\bibnamefont
  {Leuchs}},\ }\bibfield  {title} {\bibinfo {title} {Generation and direct
  detection of broadband mesoscopic polarization-squeezed vacuum},\ }\href
  {https://doi.org/10.1103/PhysRevLett.102.183602} {\bibfield  {journal}
  {\bibinfo  {journal} {Phys. Rev. Lett.}\ }\textbf {\bibinfo {volume} {102}},\
  \bibinfo {pages} {183602} (\bibinfo {year} {2009})}\BibitemShut {NoStop}%
\bibitem [{\citenamefont {Iskhakov}\ \emph {et~al.}(2012)\citenamefont
  {Iskhakov}, \citenamefont {P{\'e}rez}, \citenamefont {Spasibko},
  \citenamefont {Chekhova},\ and\ \citenamefont
  {Leuchs}}]{iskhakov2012superbunched}%
  \BibitemOpen
  \bibfield  {author} {\bibinfo {author} {\bibfnamefont {T.~S.}\ \bibnamefont
  {Iskhakov}}, \bibinfo {author} {\bibfnamefont {A.}~\bibnamefont {P{\'e}rez}},
  \bibinfo {author} {\bibfnamefont {K.~Y.}\ \bibnamefont {Spasibko}}, \bibinfo
  {author} {\bibfnamefont {M.}~\bibnamefont {Chekhova}},\ and\ \bibinfo
  {author} {\bibfnamefont {G.}~\bibnamefont {Leuchs}},\ }\bibfield  {title}
  {\bibinfo {title} {Superbunched bright squeezed vacuum state},\ }\href
  {https://doi.org/10.1364/OL.37.001919} {\bibfield  {journal} {\bibinfo
  {journal} {Opt. Lett.}\ }\textbf {\bibinfo {volume} {37}},\ \bibinfo {pages}
  {1919} (\bibinfo {year} {2012})}\BibitemShut {NoStop}%
\bibitem [{\citenamefont {Sloan}\ \emph {et~al.}(2023)\citenamefont {Sloan},
  \citenamefont {Gorlach}, \citenamefont {Tzur}, \citenamefont {Rivera},
  \citenamefont {Cohen}, \citenamefont {Kaminer},\ and\ \citenamefont
  {Soljačić}}]{sloan2023entangling}%
  \BibitemOpen
  \bibfield  {author} {\bibinfo {author} {\bibfnamefont {J.}~\bibnamefont
  {Sloan}}, \bibinfo {author} {\bibfnamefont {A.}~\bibnamefont {Gorlach}},
  \bibinfo {author} {\bibfnamefont {M.~E.}\ \bibnamefont {Tzur}}, \bibinfo
  {author} {\bibfnamefont {N.}~\bibnamefont {Rivera}}, \bibinfo {author}
  {\bibfnamefont {O.}~\bibnamefont {Cohen}}, \bibinfo {author} {\bibfnamefont
  {I.}~\bibnamefont {Kaminer}},\ and\ \bibinfo {author} {\bibfnamefont
  {M.}~\bibnamefont {Soljačić}},\ }\bibfield  {title} {\bibinfo {title}
  {Entangling extreme ultraviolet photons through strong field pair
  generation},\ }\href {https://arxiv.org/abs/2309.16466} {\bibfield  {journal}
  {\bibinfo  {journal} {arXiv preprint arXiv:2309.16466}\ } (\bibinfo {year}
  {2023})}\BibitemShut {NoStop}%
\bibitem [{\citenamefont {Grynberg}\ \emph {et~al.}(2010)\citenamefont
  {Grynberg}, \citenamefont {Aspect},\ and\ \citenamefont
  {Fabre}}]{grynberg2010introduction}%
  \BibitemOpen
  \bibfield  {author} {\bibinfo {author} {\bibfnamefont {G.}~\bibnamefont
  {Grynberg}}, \bibinfo {author} {\bibfnamefont {A.}~\bibnamefont {Aspect}},\
  and\ \bibinfo {author} {\bibfnamefont {C.}~\bibnamefont {Fabre}},\
  }\href@noop {} {\emph {\bibinfo {title} {Introduction to quantum optics: from
  the semi-classical approach to quantized light}}}\ (\bibinfo  {publisher}
  {Cambridge university press},\ \bibinfo {year} {2010})\BibitemShut {NoStop}%
\bibitem [{\citenamefont {Jackson}\ and\ \citenamefont
  {Wynne}(1982)}]{jackson82}%
  \BibitemOpen
  \bibfield  {author} {\bibinfo {author} {\bibfnamefont {D.~J.}\ \bibnamefont
  {Jackson}}\ and\ \bibinfo {author} {\bibfnamefont {J.~J.}\ \bibnamefont
  {Wynne}},\ }\bibfield  {title} {\bibinfo {title} {Interference effects
  between different optical harmonics},\ }\href
  {https://doi.org/10.1103/PhysRevLett.49.543} {\bibfield  {journal} {\bibinfo
  {journal} {Phys. Rev. Lett.}\ }\textbf {\bibinfo {volume} {49}},\ \bibinfo
  {pages} {543} (\bibinfo {year} {1982})}\BibitemShut {NoStop}%
\bibitem [{\citenamefont {Freeman}\ \emph {et~al.}(1987)\citenamefont
  {Freeman}, \citenamefont {Bucksbaum}, \citenamefont {Milchberg},
  \citenamefont {Darack}, \citenamefont {Schumacher},\ and\ \citenamefont
  {Geusic}}]{freeman1987above}%
  \BibitemOpen
  \bibfield  {author} {\bibinfo {author} {\bibfnamefont {R.}~\bibnamefont
  {Freeman}}, \bibinfo {author} {\bibfnamefont {P.}~\bibnamefont {Bucksbaum}},
  \bibinfo {author} {\bibfnamefont {H.}~\bibnamefont {Milchberg}}, \bibinfo
  {author} {\bibfnamefont {S.}~\bibnamefont {Darack}}, \bibinfo {author}
  {\bibfnamefont {D.}~\bibnamefont {Schumacher}},\ and\ \bibinfo {author}
  {\bibfnamefont {M.}~\bibnamefont {Geusic}},\ }\bibfield  {title} {\bibinfo
  {title} {Above-threshold ionization with subpicosecond laser pulses},\ }\href
  {https://doi.org/10.1103/PhysRevLett.59.1092} {\bibfield  {journal} {\bibinfo
   {journal} {Phys. Rev. Lett.}\ }\textbf {\bibinfo {volume} {59}},\ \bibinfo
  {pages} {1092} (\bibinfo {year} {1987})}\BibitemShut {NoStop}%
\bibitem [{\citenamefont {Gibson}\ \emph {et~al.}(1992)\citenamefont {Gibson},
  \citenamefont {Freeman},\ and\ \citenamefont
  {McIlrath}}]{gibson1992verification}%
  \BibitemOpen
  \bibfield  {author} {\bibinfo {author} {\bibfnamefont {G.}~\bibnamefont
  {Gibson}}, \bibinfo {author} {\bibfnamefont {R.}~\bibnamefont {Freeman}},\
  and\ \bibinfo {author} {\bibfnamefont {T.}~\bibnamefont {McIlrath}},\
  }\bibfield  {title} {\bibinfo {title} {Verification of the dominant role of
  resonant enhancement in short-pulse multiphoton ionization},\ }\href
  {https://doi.org/10.1103/PhysRevLett.69.1904} {\bibfield  {journal} {\bibinfo
   {journal} {Phys. Rev. Lett.}\ }\textbf {\bibinfo {volume} {69}},\ \bibinfo
  {pages} {1904} (\bibinfo {year} {1992})}\BibitemShut {NoStop}%
\bibitem [{\citenamefont {Hudson}(1974)}]{hudson74}%
  \BibitemOpen
  \bibfield  {author} {\bibinfo {author} {\bibfnamefont {R.}~\bibnamefont
  {Hudson}},\ }\bibfield  {title} {\bibinfo {title} {When is the wigner
  quasi-probability density non-negative?},\ }\href
  {https://doi.org/https://doi.org/10.1016/0034-4877(74)90007-X} {\bibfield
  {journal} {\bibinfo  {journal} {Rep. Math. Phys.}\ }\textbf {\bibinfo
  {volume} {6}},\ \bibinfo {pages} {249} (\bibinfo {year} {1974})}\BibitemShut
  {NoStop}%
\bibitem [{\citenamefont {Albarelli}\ \emph {et~al.}(2018)\citenamefont
  {Albarelli}, \citenamefont {Genoni}, \citenamefont {Paris},\ and\
  \citenamefont {Ferraro}}]{albarelli18}%
  \BibitemOpen
  \bibfield  {author} {\bibinfo {author} {\bibfnamefont {F.}~\bibnamefont
  {Albarelli}}, \bibinfo {author} {\bibfnamefont {M.~G.}\ \bibnamefont
  {Genoni}}, \bibinfo {author} {\bibfnamefont {M.~G.~A.}\ \bibnamefont
  {Paris}},\ and\ \bibinfo {author} {\bibfnamefont {A.}~\bibnamefont
  {Ferraro}},\ }\bibfield  {title} {\bibinfo {title} {Resource theory of
  quantum non-gaussianity and wigner negativity},\ }\href
  {https://doi.org/10.1103/PhysRevA.98.052350} {\bibfield  {journal} {\bibinfo
  {journal} {Phys. Rev. A}\ }\textbf {\bibinfo {volume} {98}},\ \bibinfo
  {pages} {052350} (\bibinfo {year} {2018})}\BibitemShut {NoStop}%
\bibitem [{\citenamefont {Law}\ \emph {et~al.}(2000)\citenamefont {Law},
  \citenamefont {Walmsley},\ and\ \citenamefont {Eberly}}]{law00}%
  \BibitemOpen
  \bibfield  {author} {\bibinfo {author} {\bibfnamefont {C.~K.}\ \bibnamefont
  {Law}}, \bibinfo {author} {\bibfnamefont {I.~A.}\ \bibnamefont {Walmsley}},\
  and\ \bibinfo {author} {\bibfnamefont {J.~H.}\ \bibnamefont {Eberly}},\
  }\bibfield  {title} {\bibinfo {title} {Continuous frequency entanglement:
  Effective finite hilbert space and entropy control},\ }\href
  {https://doi.org/10.1103/PhysRevLett.84.5304} {\bibfield  {journal} {\bibinfo
   {journal} {Phys. Rev. Lett.}\ }\textbf {\bibinfo {volume} {84}},\ \bibinfo
  {pages} {5304} (\bibinfo {year} {2000})}\BibitemShut {NoStop}%
\bibitem [{\citenamefont {Grobe}\ \emph {et~al.}(1994)\citenamefont {Grobe},
  \citenamefont {Rzazewski},\ and\ \citenamefont {Eberly}}]{grobe94}%
  \BibitemOpen
  \bibfield  {author} {\bibinfo {author} {\bibfnamefont {R.}~\bibnamefont
  {Grobe}}, \bibinfo {author} {\bibfnamefont {K.}~\bibnamefont {Rzazewski}},\
  and\ \bibinfo {author} {\bibfnamefont {J.}~\bibnamefont {Eberly}},\
  }\bibfield  {title} {\bibinfo {title} {Measure of electron-electron
  correlation in atomic physics},\ }\href
  {https://dx.doi.org/10.1088/0953-4075/27/16/001} {\bibfield  {journal}
  {\bibinfo  {journal} {J. Phys. B}\ }\textbf {\bibinfo {volume} {27}},\
  \bibinfo {pages} {L503} (\bibinfo {year} {1994})}\BibitemShut {NoStop}%
\bibitem [{\citenamefont {Brunel}(1990)}]{brunel1990harmonic}%
  \BibitemOpen
  \bibfield  {author} {\bibinfo {author} {\bibfnamefont {F.}~\bibnamefont
  {Brunel}},\ }\bibfield  {title} {\bibinfo {title} {Harmonic generation due to
  plasma effects in a gas undergoing multiphoton ionization in the
  high-intensity limit},\ }\href {https://doi.org/10.1364/JOSAB.7.000521}
  {\bibfield  {journal} {\bibinfo  {journal} {J. Opt. Soc. Am. B}\ }\textbf
  {\bibinfo {volume} {7}},\ \bibinfo {pages} {521} (\bibinfo {year}
  {1990})}\BibitemShut {NoStop}%
\bibitem [{\citenamefont {Patchkovskii}\ and\ \citenamefont
  {Muller}(2016)}]{Patchkovskii2016}%
  \BibitemOpen
  \bibfield  {author} {\bibinfo {author} {\bibfnamefont {S.}~\bibnamefont
  {Patchkovskii}}\ and\ \bibinfo {author} {\bibfnamefont {H.}~\bibnamefont
  {Muller}},\ }\bibfield  {title} {\bibinfo {title} {Simple, accurate, and
  efficient implementation of 1-electron atomic time-dependent schrödinger
  equation in spherical coordinates},\ }\href
  {https://doi.org/https://doi.org/10.1016/j.cpc.2015.10.014} {\bibfield
  {journal} {\bibinfo  {journal} {Comput. Phys. Commun.}\ }\textbf {\bibinfo
  {volume} {199}},\ \bibinfo {pages} {153} (\bibinfo {year}
  {2016})}\BibitemShut {NoStop}%
\bibitem [{\citenamefont {Schafer}\ \emph {et~al.}(1993)\citenamefont
  {Schafer}, \citenamefont {Yang}, \citenamefont {DiMauro},\ and\ \citenamefont
  {Kulander}}]{schafer1993above}%
  \BibitemOpen
  \bibfield  {author} {\bibinfo {author} {\bibfnamefont {K.}~\bibnamefont
  {Schafer}}, \bibinfo {author} {\bibfnamefont {B.}~\bibnamefont {Yang}},
  \bibinfo {author} {\bibfnamefont {L.}~\bibnamefont {DiMauro}},\ and\ \bibinfo
  {author} {\bibfnamefont {K.}~\bibnamefont {Kulander}},\ }\bibfield  {title}
  {\bibinfo {title} {Above threshold ionization beyond the high harmonic
  cutoff},\ }\href {https://doi.org/10.1103/PhysRevLett.70.1599} {\bibfield
  {journal} {\bibinfo  {journal} {Phys. Rev. Lett.}\ }\textbf {\bibinfo
  {volume} {70}},\ \bibinfo {pages} {1599} (\bibinfo {year}
  {1993})}\BibitemShut {NoStop}%
\bibitem [{\citenamefont {Corkum}(1993)}]{Corkum1994}%
  \BibitemOpen
  \bibfield  {author} {\bibinfo {author} {\bibfnamefont {P.~B.}\ \bibnamefont
  {Corkum}},\ }\bibfield  {title} {\bibinfo {title} {Plasma perspective on
  strong field multiphoton ionization},\ }\href
  {https://doi.org/10.1103/PhysRevLett.71.1994} {\bibfield  {journal} {\bibinfo
   {journal} {Phys. Rev. Lett. 71, 1994}\ } (\bibinfo {year}
  {1993})}\BibitemShut {NoStop}%
\bibitem [{\citenamefont {Lewenstein}\ \emph {et~al.}(1995)\citenamefont
  {Lewenstein}, \citenamefont {Salieres},\ and\ \citenamefont
  {L’huillier}}]{lewenstein1995phase}%
  \BibitemOpen
  \bibfield  {author} {\bibinfo {author} {\bibfnamefont {M.}~\bibnamefont
  {Lewenstein}}, \bibinfo {author} {\bibfnamefont {P.}~\bibnamefont
  {Salieres}},\ and\ \bibinfo {author} {\bibfnamefont {A.}~\bibnamefont
  {L’huillier}},\ }\bibfield  {title} {\bibinfo {title} {Phase of the atomic
  polarization in high-order harmonic generation},\ }\href
  {https://doi.org/10.1103/PhysRevA.52.4747} {\bibfield  {journal} {\bibinfo
  {journal} {Phys. Rev. A}\ }\textbf {\bibinfo {volume} {52}},\ \bibinfo
  {pages} {4747} (\bibinfo {year} {1995})}\BibitemShut {NoStop}%
\bibitem [{\citenamefont {Gaarde}\ \emph {et~al.}(1999)\citenamefont {Gaarde},
  \citenamefont {Salin}, \citenamefont {Constant}, \citenamefont {Balcou},
  \citenamefont {Schafer}, \citenamefont {Kulander},\ and\ \citenamefont
  {L’Huillier}}]{gaarde1999spatiotemporal}%
  \BibitemOpen
  \bibfield  {author} {\bibinfo {author} {\bibfnamefont {M.}~\bibnamefont
  {Gaarde}}, \bibinfo {author} {\bibfnamefont {F.}~\bibnamefont {Salin}},
  \bibinfo {author} {\bibfnamefont {E.}~\bibnamefont {Constant}}, \bibinfo
  {author} {\bibfnamefont {P.}~\bibnamefont {Balcou}}, \bibinfo {author}
  {\bibfnamefont {K.}~\bibnamefont {Schafer}}, \bibinfo {author} {\bibfnamefont
  {K.}~\bibnamefont {Kulander}},\ and\ \bibinfo {author} {\bibfnamefont
  {A.}~\bibnamefont {L’Huillier}},\ }\bibfield  {title} {\bibinfo {title}
  {Spatiotemporal separation of high harmonic radiation into two quantum path
  components},\ }\href {https://doi.org/10.1103/PhysRevA.59.1367} {\bibfield
  {journal} {\bibinfo  {journal} {Phys. Rev. A}\ }\textbf {\bibinfo {volume}
  {59}},\ \bibinfo {pages} {1367} (\bibinfo {year} {1999})}\BibitemShut
  {NoStop}%
\bibitem [{\citenamefont {Mayer}\ \emph {et~al.}(2022)\citenamefont {Mayer},
  \citenamefont {Beaulieu}, \citenamefont {Jimenez-Galan}, \citenamefont
  {Patchkovskii}, \citenamefont {Kornilov}, \citenamefont {Descamps},
  \citenamefont {Petit}, \citenamefont {Smirnova}, \citenamefont {Mairesse},\
  and\ \citenamefont {Ivanov}}]{mayer2022role}%
  \BibitemOpen
  \bibfield  {author} {\bibinfo {author} {\bibfnamefont {N.}~\bibnamefont
  {Mayer}}, \bibinfo {author} {\bibfnamefont {S.}~\bibnamefont {Beaulieu}},
  \bibinfo {author} {\bibfnamefont {A.}~\bibnamefont {Jimenez-Galan}}, \bibinfo
  {author} {\bibfnamefont {S.}~\bibnamefont {Patchkovskii}}, \bibinfo {author}
  {\bibfnamefont {O.}~\bibnamefont {Kornilov}}, \bibinfo {author}
  {\bibfnamefont {D.}~\bibnamefont {Descamps}}, \bibinfo {author}
  {\bibfnamefont {S.}~\bibnamefont {Petit}}, \bibinfo {author} {\bibfnamefont
  {O.}~\bibnamefont {Smirnova}}, \bibinfo {author} {\bibfnamefont
  {Y.}~\bibnamefont {Mairesse}},\ and\ \bibinfo {author} {\bibfnamefont
  {M.}~\bibnamefont {Ivanov}},\ }\bibfield  {title} {\bibinfo {title} {Role of
  spin-orbit coupling in high-order harmonic generation revealed by supercycle
  rydberg trajectories},\ }\href
  {https://doi.org/10.1103/PhysRevLett.129.173202} {\bibfield  {journal}
  {\bibinfo  {journal} {Phys. Rev. Lett.}\ }\textbf {\bibinfo {volume} {129}},\
  \bibinfo {pages} {173202} (\bibinfo {year} {2022})}\BibitemShut {NoStop}%
\bibitem [{\citenamefont {Woerner}\ \emph {et~al.}(2010)\citenamefont
  {Woerner}, \citenamefont {Bertrand}, \citenamefont {Kartashov}, \citenamefont
  {Corkum},\ and\ \citenamefont {Villeneuve}}]{worner2010following}%
  \BibitemOpen
  \bibfield  {author} {\bibinfo {author} {\bibfnamefont {H.}~\bibnamefont
  {Woerner}}, \bibinfo {author} {\bibfnamefont {J.}~\bibnamefont {Bertrand}},
  \bibinfo {author} {\bibfnamefont {D.}~\bibnamefont {Kartashov}}, \bibinfo
  {author} {\bibfnamefont {P.~B.}\ \bibnamefont {Corkum}},\ and\ \bibinfo
  {author} {\bibfnamefont {D.~M.}\ \bibnamefont {Villeneuve}},\ }\bibfield
  {title} {\bibinfo {title} {Following a chemical reaction using high-harmonic
  interferometry},\ }\href {https://doi.org/10.1038/nature09185} {\bibfield
  {journal} {\bibinfo  {journal} {Nature}\ }\textbf {\bibinfo {volume} {466}},\
  \bibinfo {pages} {604} (\bibinfo {year} {2010})}\BibitemShut {NoStop}%
\bibitem [{\citenamefont {Agudelo}\ \emph {et~al.}(2013)\citenamefont
  {Agudelo}, \citenamefont {Sperling},\ and\ \citenamefont
  {Vogel}}]{agudelo2013quasiprobabilities}%
  \BibitemOpen
  \bibfield  {author} {\bibinfo {author} {\bibfnamefont {E.}~\bibnamefont
  {Agudelo}}, \bibinfo {author} {\bibfnamefont {J.}~\bibnamefont {Sperling}},\
  and\ \bibinfo {author} {\bibfnamefont {W.}~\bibnamefont {Vogel}},\ }\bibfield
   {title} {\bibinfo {title} {Quasiprobabilities for multipartite quantum
  correlations of light},\ }\href
  {{https://link.aps.org/doi/10.1103/PhysRevA.87.033811}} {\bibfield  {journal}
  {\bibinfo  {journal} {Phys. Rev. A}\ }\textbf {\bibinfo {volume} {87}},\
  \bibinfo {pages} {033811} (\bibinfo {year} {2013})}\BibitemShut {NoStop}%
\bibitem [{\citenamefont {Dorfman}\ \emph {et~al.}(2016)\citenamefont
  {Dorfman}, \citenamefont {Schlawin},\ and\ \citenamefont
  {Mukamel}}]{dorfman16}%
  \BibitemOpen
  \bibfield  {author} {\bibinfo {author} {\bibfnamefont {K.~E.}\ \bibnamefont
  {Dorfman}}, \bibinfo {author} {\bibfnamefont {F.}~\bibnamefont {Schlawin}},\
  and\ \bibinfo {author} {\bibfnamefont {S.}~\bibnamefont {Mukamel}},\
  }\bibfield  {title} {\bibinfo {title} {Nonlinear optical signals and
  spectroscopy with quantum light},\ }\href
  {https://doi.org/10.1103/RevModPhys.88.045008} {\bibfield  {journal}
  {\bibinfo  {journal} {Rev. Mod. Phys.}\ }\textbf {\bibinfo {volume} {88}},\
  \bibinfo {pages} {045008} (\bibinfo {year} {2016})}\BibitemShut {NoStop}%
\bibitem [{\citenamefont {Kues}\ \emph {et~al.}(2019)\citenamefont {Kues},
  \citenamefont {Reimer}, \citenamefont {Lukens}, \citenamefont {Munro},
  \citenamefont {Weiner}, \citenamefont {Moss},\ and\ \citenamefont
  {Morandotti}}]{kues19}%
  \BibitemOpen
  \bibfield  {author} {\bibinfo {author} {\bibfnamefont {M.}~\bibnamefont
  {Kues}}, \bibinfo {author} {\bibfnamefont {C.}~\bibnamefont {Reimer}},
  \bibinfo {author} {\bibfnamefont {J.~M.}\ \bibnamefont {Lukens}}, \bibinfo
  {author} {\bibfnamefont {W.~J.}\ \bibnamefont {Munro}}, \bibinfo {author}
  {\bibfnamefont {A.~M.}\ \bibnamefont {Weiner}}, \bibinfo {author}
  {\bibfnamefont {D.~J.}\ \bibnamefont {Moss}},\ and\ \bibinfo {author}
  {\bibfnamefont {R.}~\bibnamefont {Morandotti}},\ }\bibfield  {title}
  {\bibinfo {title} {Quantum optical microcombs},\ }\href
  {https://doi.org/10.1038/s41566-019-0363-0} {\bibfield  {journal} {\bibinfo
  {journal} {Nat. Photon.}\ }\textbf {\bibinfo {volume} {13}},\ \bibinfo
  {pages} {170} (\bibinfo {year} {2019})}\BibitemShut {NoStop}%
\bibitem [{\citenamefont {Schweizer}\ \emph {et~al.}(1999)\citenamefont
  {Schweizer}, \citenamefont {Faßbinder},\ and\ \citenamefont
  {González-Férez}}]{Schweizer1999}%
  \BibitemOpen
  \bibfield  {author} {\bibinfo {author} {\bibfnamefont {W.}~\bibnamefont
  {Schweizer}}, \bibinfo {author} {\bibfnamefont {P.}~\bibnamefont
  {Faßbinder}},\ and\ \bibinfo {author} {\bibfnamefont {R.}~\bibnamefont
  {González-Férez}},\ }\bibfield  {title} {\bibinfo {title} {Model potentials
  for alkali metal atoms and li-like ions},\ }\href
  {https://doi.org/https://doi.org/10.1006/adnd.1999.0808} {\bibfield
  {journal} {\bibinfo  {journal} {At. Data Nucl. Data Tables}\ }\textbf
  {\bibinfo {volume} {72}},\ \bibinfo {pages} {33} (\bibinfo {year}
  {1999})}\BibitemShut {NoStop}%
\bibitem [{\citenamefont {Manolopoulos}(2002)}]{Manolopoulos}%
  \BibitemOpen
  \bibfield  {author} {\bibinfo {author} {\bibfnamefont {D.~E.}\ \bibnamefont
  {Manolopoulos}},\ }\bibfield  {title} {\bibinfo {title} {Derivation and
  reflection properties of a transmission-free absorbing potential},\ }\href
  {https://doi.org/10.1063/1.1517042} {\bibfield  {journal} {\bibinfo
  {journal} {J. Chem. Phys.}\ }\textbf {\bibinfo {volume} {117}},\ \bibinfo
  {pages} {9552} (\bibinfo {year} {2002})}\BibitemShut {NoStop}%
\bibitem [{\citenamefont {Kramida}\ \emph {et~al.}(2023)\citenamefont
  {Kramida}, \citenamefont {{Yu.~Ralchenko}}, \citenamefont {Reader},\ and\
  \citenamefont {{ NIST ASD Team}}}]{NIST_ASD}%
  \BibitemOpen
  \bibfield  {author} {\bibinfo {author} {\bibfnamefont {A.}~\bibnamefont
  {Kramida}}, \bibinfo {author} {\bibnamefont {{Yu.~Ralchenko}}}, \bibinfo
  {author} {\bibfnamefont {J.}~\bibnamefont {Reader}},\ and\ \bibinfo {author}
  {\bibnamefont {{ NIST ASD Team}}},\ }\href@noop {} {}\bibinfo {howpublished}
  {{NIST Atomic Spectra Database (ver. 5.11), [Online]. Available at:
  {\tt{https://physics.nist.gov/asd}} [2024, June 7]. National Institute of
  Standards and Technology, Gaithersburg, MD.}} (\bibinfo {year}
  {2023})\BibitemShut {NoStop}%
\end{thebibliography}
%

\end{document}